\documentclass[12pt]{article}
\usepackage[dvips]{graphicx}

\newcommand{\be}{\begin{eqnarray} }
\newcommand{\ee}{\end{eqnarray} }
\newcommand{\beq}{\begin{equation} }
\newcommand{\eeq}{\end{equation} }

\newcommand{\simgeq}{\; \raisebox{-0.4ex}{\tiny$\stackrel
{{\textstyle>}}{\sim}$}\;}

\newcommand{\palka}{{\Bigl |}}

\begin{document}
\begin{center}{\large\bf 
Transversity and T-odd PDFs from 
Drell-Yan processes with $pp$, $pD$ and $DD$ collisions \vskip 1cm
\rm
{
A. Sissakian\footnote{e-mail address: sisakian@jinr.ru},
O. Shevchenko\footnote{e-mail address: shev@mail.cern.ch},
A. Nagaytsev\footnote{e-mail address: nagajcev@sunse.jinr.ru},
O. Ivanov\footnote{e-mail address: ivon@jinr.ru}
 \\
{\it  Joint Institute for Nuclear Research, 141980 Dubna, Russia}\\
}
}
\end{center}
\begin{abstract}
We estimate the  single-spin  asymmetries  (SSA) 
which provide the access to transversity as well as to Boer-Mulders and Sivers 
PDFs via investigation of the single-polarized Drell-Yan (DY) processes
with $ pp$, $pD$ and $DD$ collisions 
available to RHIC, NICA, COMPASS,  and J-PARC. 
The feasibility of these SSA is studied with the new generator of polarized DY events.
The performed  estimations demonstrate
that there exist the such kinematical regions where SSA are 
presumably measurable. 
The most useful for PDFs extraction are the limiting kinematical 
ranges, where one can neglect the sea PDFs contributions which  occur at large values of Bjorken x. 
It is of interest that on the contrary to the Sivers PDF, the  transversity PDF
is presumably accessible only in the especial kinematical region.
On the contrary to the option with the symmetric collider mode (RHIC, NICA),
this is of 
importance for the COMPASS experiment and the future J-PARC facility where the 
fixed target mode is available.

\end{abstract}
\begin{flushleft}
\end{flushleft}

\section{Introduction}

In this paper we focus on DY processes in collisions of transversely polarized protons
and deutrons, providing us an access to the very important and still
poorly known sea and valence transversity, Boer-Mulders and Sivers PDFs
in proton.  
At present the such processes are available to RHIC \cite{rhic}.   
Their studies are planned at J-PARC \cite{sawada} facility and, in principle, are possible\footnote{At present 
the Drell-Yan program at COMPASS focuses on the pion-proton (deutron) collisions (see, for example,
\cite{compass_dy}). However, the possibility to study 
DY processes  with $pp$ and $pd$ collisions is now also under discussion. } at
COMPASS, where unpolarized proton beam and both polarized proton and deutron targets 
are available. Besides,
at present JINR starts the new NICA/MPD project based on the development of the
existing Nuclotron accelerator for the new facility creation:
the heavy and light nucleus collider NICA \cite{nica}, \cite{nica1}. 
In particular, the possibility is now considered to study the  collisions of the polarized proton and deutron beams 
at the second interaction point (IP) at NICA.
It is of importance, that all these experiments do not duplicate each other, 
but are complementary, providing us by the
the information on the different PDFs measured in the different\footnote{
For $pp$ collisions the center of mass energy is 200 GeV for RHIC.
For COMPASS it can be varied from 20 to 27 GeV (upper bound corresponds to
400 GeV primary proton SPS beam). For J-PARC $\sqrt{s}$ is planned to be
about 8 GeV at first  stage and about 10 GeV for the second stage. 
For NICA it is planned to be about 20-26 GeV.
} kinematical regions.


The leading twist $k_T$ integrated transversity PDF $\Delta_{T}q \equiv h_{1q} $, 
as well as the leading twist
unpolarized $q\equiv f_{1q}$ and longitudinally polarized (helicity) $\Delta q
\equiv g_{1q}$ PDFs, is of the crucial importance for understanding of the 
nucleon spin structure (see Ref. \cite{review} for the comprehensive  review).
At the same time, nowadays the study of quark transverse momentum $k_T$ dependent
PDFs is also among the special issues in
hadron physics. Of particular interest, are two
leading-twist T-odd $k_T$ dependent PDFs: 
Sivers function $f_{1T}^{\perp q}(x, k_T^2)$ and 
Boer-Mulders function $h_{1q}^{\perp}(x, k_T^2)$.
While Sivers function represents the unpolarized parton
distribution  in a transversely polarized hadron, the
Boer-Mulders function denotes the parton transversity distribution
in the unpolarized hadron. 

At present 
the Boer-Mulders PDF is still not measured, 
while 
the Sivers \cite{efremov_old, efremov_new} 
and transversity \cite{ansel_kotz} 
PDFs 
were (preliminarily, with rather big uncertainties)
extracted from  the SIDIS data
collected by HERMES 
\cite{hermes} 
and COMPASS \cite{compass} collaborations. 
At the same time 
the analysis of SIDIS data suffers from the poor knowledge of the fragmentation
functions, and especially it concerns the Collins fragmentation function
which is necessary for the transversity extraction \cite{ansel_kotz}.  
In this respect the 
Drell-Yan processes possess the essential advantage since  
they are free of any fragmentation functions.
Besides, DY measurements {\it should} accompany SIDIS measurements  
to check the important QCD prediction \cite{siv_collins} (see also
\cite{collins_doklad} and references therein) 
\be
\label{sign}
f_{1T}^{\perp}\palka_{DY}=-f_{1T}^{\perp}\palka_{SIDIS},\quad 
h_{1}^{\perp}\palka_{DY}=-h_{1}^{\perp}\palka_{SIDIS}
\ee
for T-odd PDFs $f_{1T}^{\perp}$  and $h_{1}^{\perp}$. It is relevant to notice in this connection
that while the Sivers PDF
was already extracted from SIDIS (with rather bad precision but at least the sign of $f_{1T}^\perp\palka_{SIDIS}$
is seen) there still was not the respective analysis for $h_1^\perp\palka_{SIDIS}$. 
Recently in the paper \cite{barone_bm_model}
the estimations were presented of the possibility to extract the azimuthal asymmetry  $\langle \cos2\phi\rangle$
(giving the access to $h_1^\perp$) from the combined analysis of all existing and planned SIDIS measurements.
Thus, if this SIDIS program would be realized, then DY measurements can allow to check sign change effect 
for  both Sivers and Boer-Mulders PDFs.

It is well known that  the double transversely polarized DY process
$H_1^{\uparrow}H_2^{\uparrow} \to l^{+}l^{-}X$  allows  
to directly extract the transversity distributions 
(see Ref. \cite{review} for review).  
However, at the same number of collected DY events,
the double spin asymmetries suffer from the much more large statistical errors
(product of the beam and target/beam polarizations in the denominator) than the single spin asymmetries.
This is especially important for the double-polarized 
DY processes with $pp$ (as well as with $pD$ and $DD$) collisions,   
where because of the small value of transversity PDF for the sea antiquark
in proton (neutron), the double spin asymmetry is estimated to be a few percents
{\it maximum} \cite{vogelsang}.
Thus, in the case of  $pp$, $pD$ and $DD$ collisions we focus on here, 
the double polarized DY processes seem to be not too useful
and we need 
an alternative possibility allowing 
to extract the transversity PDF from the combined analysis of unpolarized 
\be
\label{unpolDY}
H_1  H_2 \to l^{+}l^{-}\, X, 
\ee
and single-polarized
\be
\label{polDY}
H_1  H_2^{\uparrow} \to l^{+}l^{-}\, X 
\ee
DY processes.  
Besides, namely unpolarized and single-polarized DY processes give
us also an access to Boers-Mulders and Sivers PDFs, which are very
intriguing and interesting objects in themselves.
On the other hand, in the processes (\ref{unpolDY}) and (\ref{polDY})  
the access to PDFs we are interesting in is rather difficult since 
they enter the respective cross-sections \cite{bmodel}  
in the complex convolution 
with each other, so that at first sight it is impossible to avoid some
models on the $k_T$ dependence of PDFs.  
To solve this problem
the $q_T$ weighting approach \cite{kotz,mul1,mul2,mul3} was recently applied 
in Ref. \cite{efremov_old} to Sivers effect in the single-polarized
DY processes (\ref{polDY}), and in Refs. \cite{approach,our2} with respect to
transversity and Boer-Mulders PDF extraction from both unpolarized and single-polarized
DY processes (\ref{unpolDY}) and (\ref{polDY}).  
In two last cited papers we considered the  DY processes 
with 
antiproton-proton and pion-proton collisions. 
At the same time the DY processes with the proton (deutron)-proton(deutron) collisions are also
very important since they provide the access to sea PDFs. 
Within this paper we will estimate both types of single-spin asymmetries (SSA), which 
give us respectively access to Sivers PDF \cite{efremov_old,efremov_new} and to
transversity and Boer-Mulders PDFs \cite{approach,our2}.
At first sight it seems that DY processes with proton(deutron)-proton(deutron) collisions
are strongly suppressed because there is no valence antiquark in the initial 
state there. However, we will see
that there exist the kinematical regions where both SSA take quite considerable
values.


\section{Transversity and T-odd PDFs via Drell-Yan processes with $pp$ collisions}

The procedure proposed in Refs. \cite{approach,our2} allows us
to extract from the processes (\ref{unpolDY})
and (\ref{polDY})
the transversity $h_1$ and the
first moment 
\be
\label{hperpmom}
h_{1q}^{\perp(1)}(x)\equiv\int d^2{\bf k}_T\left(\frac{{\bf k}_T^2}{2M_p^2}\right)
h_{1q}^\perp(x_p,{\bf k}_T^2)
\ee
of Boer-Mulders $h_1^{\perp(1)}$ PDF directly, without any model assumptions 
about $k_T$-dependence 
of $h_1^\perp(x,k_T^2)$.
Applied to unpolarized DY process  
(\ref{unpolDY}) with $pp$ collisions
this general procedure gives\footnote{Eq. (\ref{e1}) is obtained within the  quark parton model. It is of importance that 
the large values of coefficient at $\cos2\phi$ in the ratio of DY cross-sections
can not be explained by the leading and next-to-leading order perturbative QCD
corrections as well as by the high twists effects (see \cite{bmodel} and references therein).} 
\be
\label{e1}
\hat k\palka_{pp\rightarrow l^+l^-X} =
8\frac{\sum_qe_q^2[\bar h_{1q}^{\perp(1)}(x_1) h_{1q}^{\perp(1)}(x_2)
+(q\rightarrow \bar q)]}
{\sum_qe_q^2[\bar f_{1q}(x_1)f_{1q}(x_2)+(q\rightarrow \bar q)]},
\ee
where $\hat k$ is the coefficient at $\cos 2\phi$ dependent part of the properly $q_{T}$ weighted  
ratio of unpolarized cross-sections:
\be
\label{r1}
& & \hat R=\frac{\int d^2 {\bf q}_T [{\bf |}{\bf q}_T{\bf |}^2/{M^2_p}][d\sigma^{(0)}/d\Omega]}{\int d^2{\bf q}_T\sigma^{(0)}},\\
\label{r2}
 &  & \hat R=\frac{3}{16\pi}(\gamma(1+\cos^2\theta)+\hat k\cos 2\phi\sin^2\theta).
\ee
At the same time, in the case of single-polarized DY process  
(\ref{polDY}),
operating just as in Ref. \cite{approach},
one gets
\be
\label{ssa_bm}
\hat A_{h}  & = &
-\frac{1}{2} \frac{\sum_q e_q^2[
\bar h_{1q}^{\perp(1)}(x_p) h_{1q}(x_{p^\uparrow})+(q \rightarrow \bar q)]}
{\sum_q e_q^2 [\bar f_{1q}(x_p) f_{1q}(x_{p^\uparrow})+(q \rightarrow \bar q)]},
\ee
where the single spin asymmetry (SSA) $\hat A_h$ is defined as
\be
\label{ssa_bm_angle}
\hat A_{h}=\frac{\int d\Omega d\phi_{S_2}\int d^2{\bf q}_T(|{\bf q}_T|/M_p)\sin(\phi+\phi_{S_2})[d\sigma({\bf S}_{2T})-d\sigma(-{\bf S}_{2T})]}{\int d\Omega d \phi_{S_2}\int d^2{\bf q}_T[d\sigma({\bf S}_{2T})+d\sigma(-{\bf S}_{2T})]}.
\ee
All the notations used are the same as in Ref. \cite{approach} (see Ref. \cite{review} for details on kinematics 
in the Collins-Soper frame we deal with). 

Notice that SSA $\hat A_h$ is analogous to  asymmetry 
$A_{UT}^{\sin(\phi-\phi_S)\frac{q_T}{M_N}}$
(weighted with $\sin(\phi-\phi_S$) and the same weight
$q_T/M_N$) applied in Ref. \cite{efremov_old} with respect to 
the Sivers effect investigation in the single-polarized DY processes. 
For DY process $pp^\uparrow\rightarrow l^+l^-X$ we study here
the expressions for $A_{UT}^{\sin(\phi-\phi_S)\frac{q_T}{M_N}}$ look
as (see Eqs. (14), (15) in Ref. \cite{efremov_old} )   
\be
\label{ssa_siv_angle}
A_{UT}^{\sin(\phi-\phi_S)\frac{q_T}{M_N}}=\frac{\int d\Omega d\phi_{S_2}
\int d^2{\bf q}_T(|{\bf q}_T|/M_p)\sin(\phi-
\phi_{S_2})[d\sigma({\bf S}_{2T})-d\sigma(-{\bf S}_{2T})]}{ \frac{1}{2}\int 
d\Omega d \phi_{S_2}\int d^2{\bf q}_T[d\sigma({\bf S}_{2T})+d\sigma(-{\bf S}_{2T})]},
\ee
\be
\label{ssa_siv}
A_{UT}^{\sin(\phi-\phi_S)\frac{q_T}{M_N}}  & = &
2 \frac{\sum_q e_q^2[
\bar f_{1T}^{\perp(1)q}(x_{p^\uparrow}) f_{1q}(x_p)+(q \rightarrow \bar q)]}
{\sum_q e_q^2 [\bar f_{1q}(x_{p^\uparrow}) f_{1q}(x_p)+(q \rightarrow \bar q)]},
\ee
where
\be
\label{siversmom}
f_{1T}^{\perp(1)q}(x)\equiv\int d^2{\bf k}_T\left(\frac{{\bf k}_T^2}{2M_p^2}\right)
f_{1T}^{\perp q}(x,{\bf k}_T^2)
\ee
is the first moment of the Sivers function
$f_{1T}^{\perp q}(x,{\bf k}_T^2)$.
Notice that factor $1/2$ in denominator of Eq. (\ref{ssa_siv_angle}) 
(see also Eq. (7) in Ref. \cite{efremov_rhic} )
was introduced in Ref. \cite{efremov_old} (where the Sivers
effect was studied in both SIDIS and DY processes)  
for consistence with the respective semi-inclusive SSA studied by the HERMES
-- see \cite{efremov_new}  and references therein.
Since within this paper we also will study SSA given by Eqs. (\ref{ssa_siv_angle})
and (\ref{ssa_siv}), for comparison purposes  it is convenient to introduce\footnote{
Certainly, from the practical point of view, the such rescaling is not 
especially useful:
when the asymmetry is multiplied by a number the error is  multiplied by the
same number too. However, it is convenient to consider both studied SSA
given in the same scale in order to estimate and compare their feasibility at the same 
statistics of DY events.}, by analogy, SSA 
 \be
\label{rescaled}
A_{UT}^{\sin(\phi+\phi_S)\frac{q_T}{M_N}}=2 \hat A_h 
=- \frac{\sum_q e_q^2[
\bar h_{1q}^{\perp(1)}(x_p) h_{1q}(x_{p^\uparrow})+(q \rightarrow \bar q)]}
{\sum_q e_q^2 [\bar f_{1q}(x_p) f_{1q}(x_{p^\uparrow})+(q \rightarrow \bar q)]}.
\ee

Let us now consider SSA given by Eqs. (\ref{ssa_siv}) and (\ref{rescaled}). 
On the contrary to valence PDFs, the sea PDFs dominate at small $x$ and rapidly
die out when $x$ increases. That is why in the case of $pp^\uparrow $ collisions 
we deal with
the regions are of importance where the Bjorken $x$ for sea PDFs take the small values,
while, by virtue of the relation 
\be
x_px_{p^\uparrow} = Q^2/s,
\ee
the valence PDFs occur at large $x$ values. Indeed, in such the regions we can neglect
the contributions to SSA containing sea PDFs at large $x$ 
(later we will see that this is really good approximation)
and, thereby, to essentially cancel the number of extra unknown PDFs 
entering the asymmetries.   

Thus, let us consider
two limiting cases $x_p  \gg x_{p^\uparrow} $ and $x_p \ll x_{p^\uparrow} $.
 In the first case
\be
 x_{unpol} \gg x_{pol}, 
\label{limit1}
\ee
neglecting\footnote{Notice that all over the paper we neglect
in our calculations the contributions of strange PDFs which produce 
really tiny corrections.} the terms containing the sea PDFs at large $x_p$,
one arrives at the simplified equations  
\be
\label{siv_asympt_01}
A_{UT}^{\sin(\phi-\phi_S)\frac{q_T}{M_N}}{\Bigl |}_{x_p\gg x_{p^\uparrow}}  \simeq  
2 \frac{
4\bar f_{1T}^{\perp(1)u}(x_{p^\uparrow}) f_{1u}(x_p) + 
\bar f_{1T}^{\perp(1)d}(x_{p^\uparrow}) f_{1d}(x_p)   }
{4\bar f_{1u}(x_{p^\uparrow}) f_{1u}(x_p)+\bar f_{1d}(x_{p^\uparrow}) f_{1d}(x_p)   },
\ee
and
\be
\label{bm_asympt_01}
A_{UT}^{\sin(\phi+\phi_S)\frac{q_T}{M_N}}{\Bigl |}_{x_p\gg x_{p^\uparrow}}  \simeq
- \frac{4
h_{1u}^{\perp(1)}(x_p) \bar h_{1u}(x_{p^\uparrow})+
h_{1d}^{\perp(1)}(x_p) \bar h_{1d}(x_{p^\uparrow})  }
{4f_{1u}(x_p) \bar f_{1u}(x_{p^\uparrow}) +f_{1d}(x_p) \bar f_{1d}(x_{p^\uparrow})   }.
\ee
Then, taking into account the quark charges 
and $u$ quark dominance at large $x$, Eqs. (\ref{siv_asympt_01}) and (\ref{bm_asympt_01})  
are essentially given by
\be
\label{siv_asympt_1}
A_{UT}^{\sin(\phi-\phi_S)\frac{q_T}{M_N}}{\Bigl |}_{x_p\gg x_{p^\uparrow}}  \simeq  
2 \frac{
\bar f_{1T}^{\perp(1)u}(x_{p^\uparrow}) f_{1u}(x_p)}
{\bar f_{1u}(x_{p^\uparrow}) f_{1u}(x_p)}
= 2 \frac{
\bar f_{1T}^{\perp(1)u}(x_{p^\uparrow})}
{\bar f_{1u}(x_{p^\uparrow})},
\ee
and
\be
\label{bm_asympt_1}
A_{UT}^{\sin(\phi+\phi_S)\frac{q_T}{M_N}}{\Bigl |}_{x_p\gg x_{p^\uparrow}}  \simeq
- \frac{
h_{1u}^{\perp(1)}(x_p) \bar h_{1u}(x_{p^\uparrow})}
{f_{1u}(x_p) \bar f_{1u}(x_{p^\uparrow})}.
\ee
Analogously, in the second limiting case
\be
\label{limit2}
x_{unpol} \ll x_{pol}, 
\ee
one gets
\be
\label{siv_asympt_02}
A_{UT}^{\sin(\phi-\phi_S)\frac{q_T}{M_N}}{\Bigl |}_{x_p\ll x_{p^\uparrow}}  \simeq  
2 \frac{
4 f_{1T}^{\perp(1)u}(x_{p^\uparrow}) \bar f_{1u}(x_p) + 
f_{1T}^{\perp(1)d}(x_{p^\uparrow}) \bar f_{1d}(x_p)   }
{4 f_{1u}(x_{p^\uparrow})\bar f_{1u}(x_p)+ f_{1d}(x_{p^\uparrow}) \bar f_{1d}(x_p)   },
\ee
and
\be
\label{bm_asympt_02}
A_{UT}^{\sin(\phi+\phi_S)\frac{q_T}{M_N}}{\Bigl |}_{x_p\ll x_{p^\uparrow}}  \simeq
- \frac{4
\bar h_{1u}^{\perp(1)}(x_p)  h_{1u}(x_{p^\uparrow})+
\bar h_{1d}^{\perp(1)}(x_p)  h_{1d}(x_{p^\uparrow})  }
{4 \bar f_{1u}(x_p) f_{1u}(x_{p^\uparrow}) +\bar f_{1d}(x_p) f_{1d}(x_{p^\uparrow})   },
\ee
with $d$ quark contributions,  while
\be
\label{siv_asympt_2}
A_{UT}^{\sin(\phi-\phi_S)\frac{q_T}{M_N}}{\Bigl |}_{x_p\ll x_{p^\uparrow}}  \simeq  
2 \frac{
 f_{1T}^{\perp(1)u}(x_{p^\uparrow}) \bar f_{1u}(x_p)}
{ f_{1u}(x_{p^\uparrow}) \bar f_{1u}(x_p)}
= 2 \frac{
 f_{1T}^{\perp(1)u}(x_{p^\uparrow})}
{ f_{1u}(x_{p^\uparrow})},
\ee
and
\be
\label{bm_asympt_2}
A_{UT}^{\sin(\phi+\phi_S)\frac{q_T}{M_N}}{\Bigl |}_{x_p\ll x_{p^\uparrow}}  \simeq
- \frac{
\bar h_{1u}^{\perp(1)}(x_p)  h_{1u}(x_{p^\uparrow})}
{\bar f_{1u}(x_p) f_{1u}(x_{p^\uparrow})},
\ee
neglecting $d$ quark contribution.

Later (see Section 3) 
we will see that 
even the double approximations given by  Eqs. 
(\ref{bm_asympt_1}) and (\ref{bm_asympt_2})
are in a very good agreement with Eqs. (\ref{ssa_siv}) and (\ref{rescaled}).
It is of importance
since allows us to cancel out the extra unknown variables
entering the equations for measured asymmetries.
In particular, this give us the interesting 
possibility to extract the ratios 
$h_{1u}/h_{1u}^{\perp(1)}$  and $\bar h_{1u}/\bar h_{1u}^{\perp(1)}$ 
directly, without application of the 
fitting procedure with a set of assumptions on extra unknown variables. Indeed, let us return
to the  unpolarized DY process with $pp$ collisions, Eqs. (\ref{e1})-(\ref{r2}).  
In the limiting cases $x_1 \gg x_2$ and $x_1 \ll x_2$ Eq. (\ref{e1}) is reduced to the equations 
\be
\label{unpol1}
\hat k \palka_{x_1\gg x_2}\simeq 8\frac{h_{1u}^{\perp(1)}(x_1)\bar h_{1u}^{\perp(1)}(x_2)}{f_{1u}(x_1)\bar f_{1u}(x_2)},
\ee
and
\be
\label{unpol2}
\hat k \palka_{x_1\ll x_2}\simeq 8\frac{\bar h_{1u}^{\perp(1)}(x_1) h_{1u}^{\perp(1)}(x_2)}{\bar f_{1u}(x_1) f_{1u}(x_2)},\ee
respectively.
Then, having in our disposal the quantities  $\hat k(x_1,x_2)$ 
and $\hat A_{UT}^{\sin(\phi+\phi_S)\frac{q_T}{M_N}}(x_p =x_1,x_{p^\uparrow}=x_2)$
measured in unpolarized and single-polarized DY processes, and, combining Eqs. (\ref{bm_asympt_1}),
(\ref{bm_asympt_2}) with Eqs. (\ref{unpol1}), (\ref{unpol2}),  
we can obtain the ratios 
$h_{1u}/h_{1u}^{\perp(1)}$  and $\bar h_{1u}/\bar h_{1u}^{\perp(1)}$ 
from the equations 
\be
\frac{\bar h_{1u}(x_1)}{\bar h_{1u}^{\perp(1)}(x_1)}
\simeq-8\frac{\hat A_{UT}^{\sin(\phi+\phi_S)\frac{q_T}{M_N}}}{\hat k}\palka_{x_1\gg x_2}, \quad
\frac{h_{1u}(x_1)}{h_{1u}^{\perp(1)}(x_1)}
\simeq-8\frac{\hat A_{UT}^{\sin(\phi+\phi_S)\frac{q_T}{M_N}}}{\hat k}\palka_{x_1\ll x_2}.
\ee

\section{Estimations on SSA in $pp$ collisions}
\begin{figure}[h!]
        \begin{tabular}{cc}
\includegraphics[height=5cm]{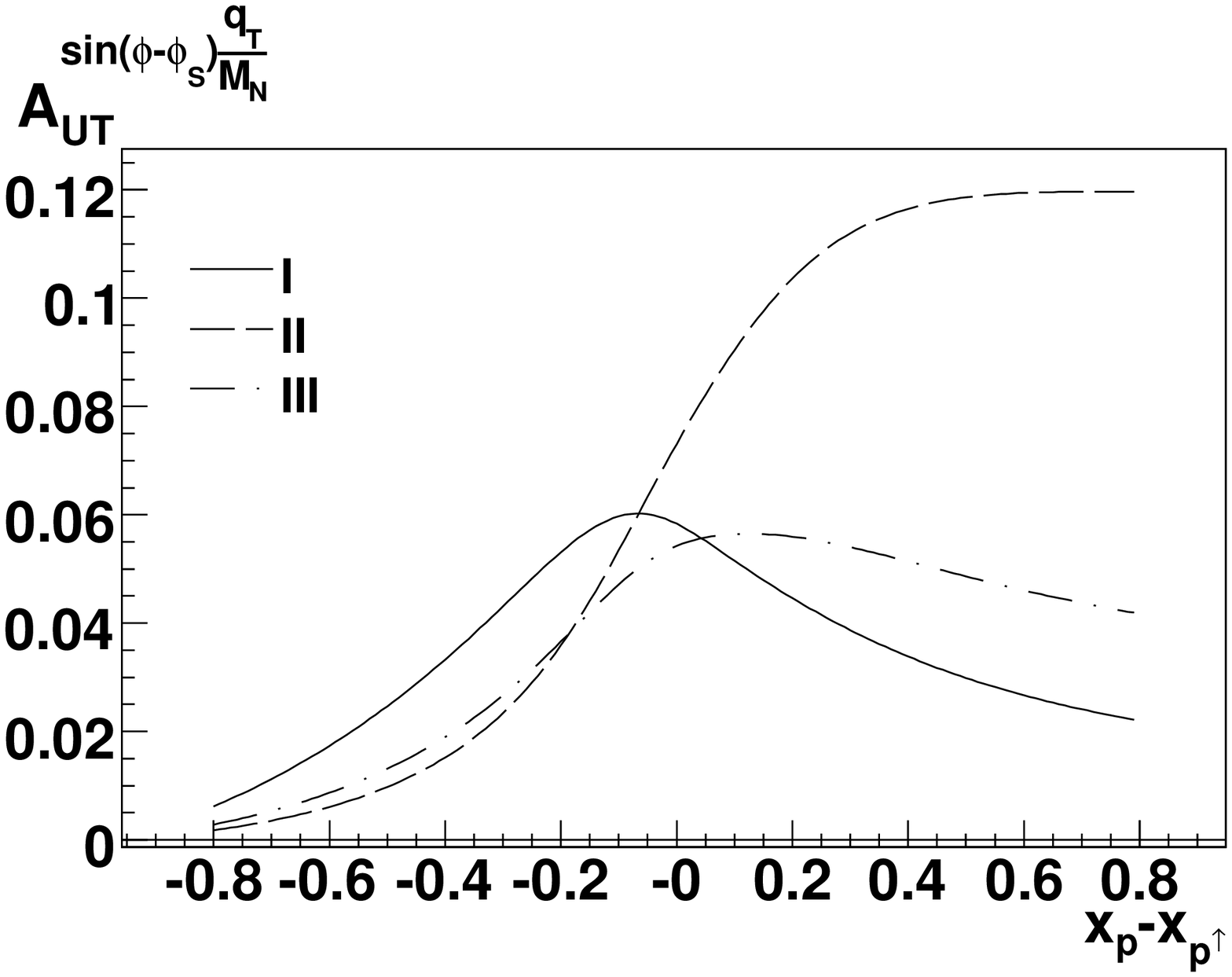} &\includegraphics[height=5cm]{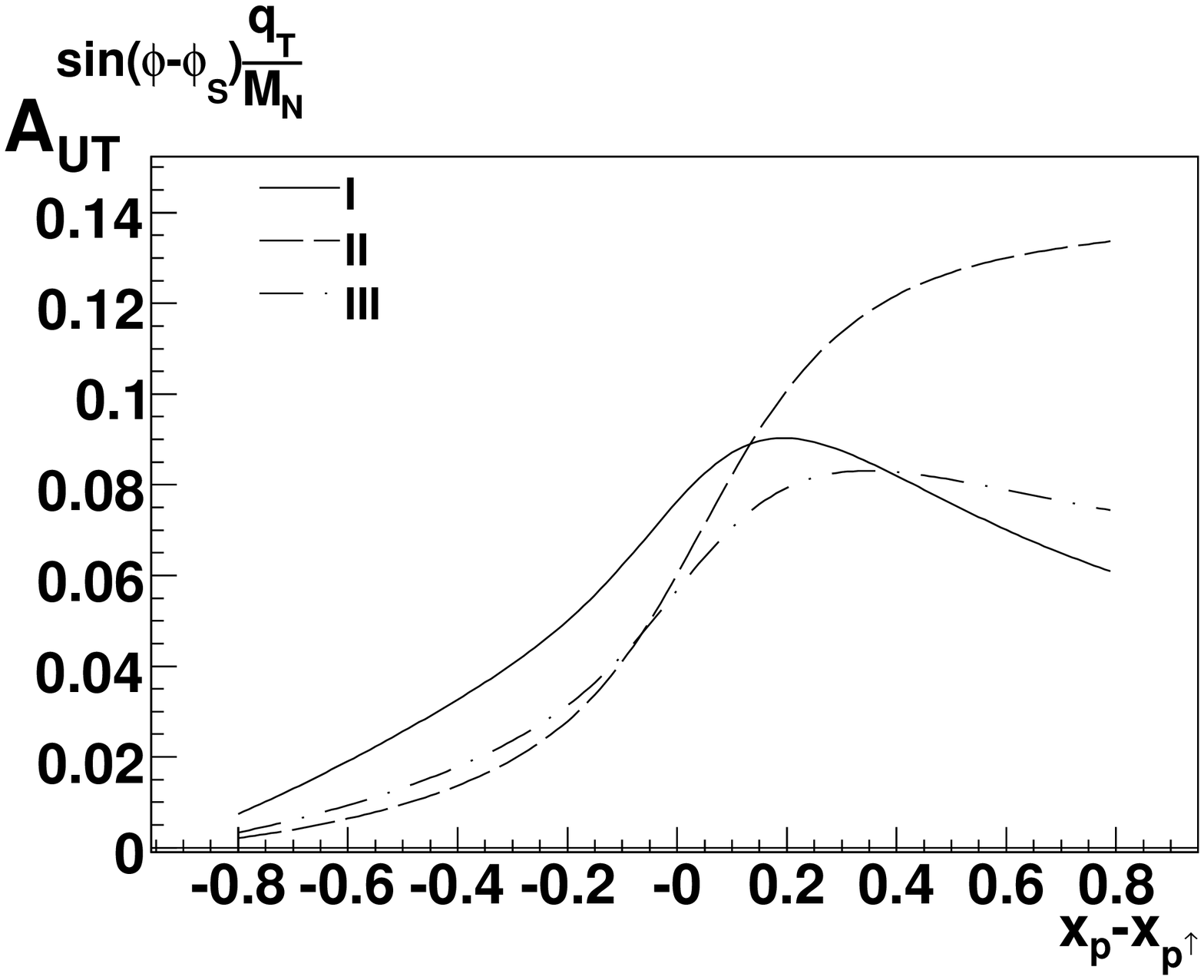} 
        \end{tabular}
        \caption{
        Estimation of SSA $A_{UT}^{\sin(\phi-\phi_S)\frac{q_T}{M_N}}\palka_{pp^\uparrow}$
   for NICA, s=400$GeV^2$,
        with $Q^2=4\,GeV^2$ (left) and $Q^2=15\,GeV^2$ (right). 
        Rome numbers $I,II$
        denote respectively fits I and II from  Ref. \cite{efremov_old} and $III$ denotes
the fit from  Ref.  \cite{efremov_new}.
        }
 \label{fig:sivers1}
\end{figure}

Let us first estimate SSA 
$A_{UT}^{\sin(\phi-\phi_S)\frac{q_T}{M_N}}\palka_{pp^\uparrow \rightarrow l^+l^-X}$ given by Eq. (\ref{ssa_siv}) .
Notice that for the RHIC kinematics SSA $A_{UT}^{\sin(\phi-\phi_S)}$ 
weighted only with the projecting factor $\sin(\phi-\phi_S)$  
was already in detail studied 
in Ref. \cite{efremov_rhic}. This asymmetry
within the Gaussian model applied in \cite{efremov_rhic} is just
proportional to the $q_T$ weighted SSA   $A_{UT}^{\sin(\phi-\phi_S)\frac{q_T}{M_N}}$
which we consider in this paper:
$
A_{UT}^{\sin(\phi-\phi_S)} = a_{Gauss}^{DY}\,A_{UT}^{\sin(\phi-\phi_S)\frac{q_T}{M_N}}
$
($ a_{Gauss}^{DY} \simeq 0.81\cdot(1\pm10\%)$), so that we need not to repeat here the
calculations for the RHIC kinematical domain.
Instead, we present the estimations for the NICA kinematics, 
where two  colliding  $10 \sim 13\,GeV$ proton beams 
assumed to be available \cite{nica}, \cite{nica1}.
We perform the calculations  for $Q^2$ values below  and above  
$J/\Psi$ threshold $Q^2=9.5\,GeV^2$.
For the estimations we use three different fits for the Sivers function: 
fits I and II from Ref. \cite{efremov_old}
and also the latest fit from Ref. \cite{efremov_new}, which we denote as fit III. 
These fits are given by the following parametrizations
\be
\label{ef1}
 &  & \mbox{\rm Fit I:}\quad xf_{1T}^{\perp(1)u}=-xf_{1T}^{\perp(1)d}=0.4x(1-x)^5,\\
 &  & \mbox{\rm Fit II:}\quad xf_{1T}^{\perp(1)u}=-xf_{1T}^{\perp(1)d}=0.1x^{0.3}(1-x)^5,\\
 \label{ef3} 
&  & \mbox{\rm Fit III:}\quad xf_{1T}^{\perp(1)u}=
-xf_{1T}^{\perp(1)d}=(0.17...0.18)x^{0.66}(1-x)^5.
\ee
For the first moments (\ref{siversmom}) of the sea Sivers PDFs entering Eq. (\ref{ssa_siv})
we use the model (with the positive sign) proposed in Ref. \cite{efremov_rhic} (see Eqs. (10) and (11) in
Ref. \cite{efremov_rhic}):
\be
\label{seamodel1}
\frac{f_{1T}^{\perp(1)\bar q}(x)}{f_{1T}^{\perp(1)q}(x)}=  
\frac{f_{1\bar u}(x) +f_{1\bar d}(x)}{f_{1u}(x)+ f_{1 d}(x)}
\ee
For the unpolarized PDFs entering Eq. (\ref{ssa_siv}) we use GRV94 \cite{grv94} parametrization.
The results of estimations for the different $Q^2$ values are presented in  
Fig. \ref{fig:sivers1}.

Looking at Fig. \ref{fig:sivers1}  one can see that the 
asymmetry takes the largest values near zero value of $x_p-x_{p^\uparrow}$ 
and when this difference becomes positive.

Notice that besides of parametrizations I, II and III on the Sivers PDF
there exist also the parametrizations from Refs. \cite{anselmino05}, \cite{vogelsang05}.
Our estimations\footnote{Dealing with the parametrization from 
Ref. \cite{anselmino05} one should remember that 
the notations for Sivers PDF $\Delta^N f_{q/H^\uparrow}$ and $f_{1T}^\perp$ differ by the sign and extra
multiplier:
$f_{1T}^\perp(x,{\bf k}_T^2)=-(M/2|{\bf k}_T|)\Delta^N f_{q/H^\uparrow}(x,{\bf k}_T^2)$ (see ``Trento conventions'' in
Ref. \cite{trento}).} with these parametrizations show that the values 
of $A_{UT}^{\sin(\phi-\phi_S)}$ are quite similar to the respective 
values obtained with the parametrizations I, II, III in the region $x_p>x_{p^\uparrow}$
and even higher in the region  $x_p<x_{p^\uparrow}$. Thus, the prediction on
$A_{UT}^{\sin(\phi-\phi_S)}$ with parametrizations I, II, III could be considered
as even underestimated predictions: if  
$A_{UT}^{\sin(\phi-\phi_S)}$ predicted by parametrizations I, II, III  will be seen within the errors,
then the larger SSA predicted by the fits from Refs. \cite{anselmino05} and \cite{vogelsang05}
will be definitely measurable too.

Our calculations performed for the COMPASS and J-PARC kinematics 
(in comparison with RHIC rather close to the NICA one) 
produce for SSA $A_{UT}^{\sin(\phi-\phi_S)\frac{q_T}{M_N}}$
the plots  very similar to the respective plots for the NICA kinematical range.
The such behavior of $A_{UT}^{\sin(\phi-\phi_S)\frac{q_T}{M_N}}$ encourage one  that it can be measured not only 
in the collider mode (RHIC, NICA) but also in the fixed target experiments
(COMPASS, J-PARC). The point is
that, as a rule, in the experiments with a fixed target the acceptance of the detector
allows us to register mainly the events with the positive $x_F$:
\be 
\label{restriction} 
x_F \equiv x_{beam} - x_{target} \simgeq 0, 
\ee
where $x_{beam}$ and $x_{target}$ are the Bjorken $x$ values for the quarks 
inside of the beam and target protons, respectively.  Thus, the option 
$x_p>x_{p^\uparrow}$ can be realized by both COMPASS and J-PARC facilities,
where unpolarized proton beam and polarized proton target can be available.
%
%
%
On the hand, the region $x_p<x_{p^\uparrow}$, where SSA $A_{UT}^{\sin(\phi-\phi_S)\frac{q_T}{M_N}}$
is also quite considerable, definitely can be reached by the collider experiments RHIC and NICA,
and, presumably, by J-PARC, where the option with the polarized beam 
is also planned (see Ref. \cite{sawada}).

Let us now estimate SSA $A_{UT}^{\sin(\phi+\phi_S)\frac{q_T}{M_N}}$ 
given by Eq. (\ref{rescaled}). 
Since neither the Boer-Mulders function 
nor its first moment
are still not measured, we will use in our calculation
 the Boer's model (Eq. (50) in Ref. \cite{bmodel}) which produces
the good fit for the  NA10 \cite{NA10} and E615 \cite{E615} data
on (anomalously large) $\cos(2\phi)$  dependence of DY cross-sections.
This model gives for the first moment (\ref{hperpmom}) entering Eq. (\ref{rescaled})  
the value $ h_{1q}^{\perp(1)}(x)\simeq0.163f_1(x)$.
We also apply the following assumption for the first moment of the sea Boer-Mulders PDF
\be
\label{seamodel2}
\frac{h_{1\bar q}^{\perp(1)}(x)}{h_{1q}^{\perp(1)}(x)}=\frac{f_{1\bar q}(x)}{f_{1 q}(x)}.
\ee
Notice that this assumptions is similar to the assumption (\ref{seamodel1}), however,
namely Eq. (\ref{seamodel2}) is consistent with the Boer's model which we use here.

\begin{figure}[h!]
        \includegraphics[height=5cm]{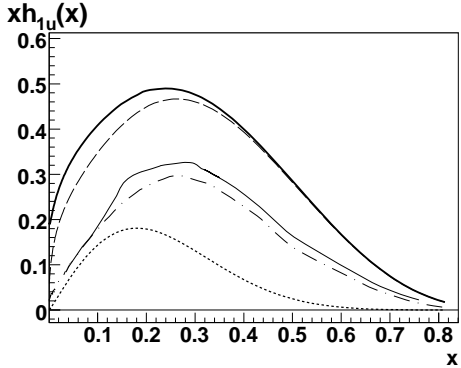}
        \caption{Results of Ref. \cite{ansel_kotz} on $h_{1u}$ in comparison with the results
        obtained with two versions of evolution model and with the Soffer bound. 
        The bold solid line (top) corresponds to upper bound given by the Soffer inequality.
        Dashed line corresponds to the evolution model with the Soffer inequality saturation
        at the initial model scale $Q_0^2=0.23\,GeV^2$.
        The solid line corresponds to the upper boundary of the error band on $h_{1u}$.
        The dashed-dotted line corresponds to the evolution model, where 
        $h_{1u,\bar u}=\Delta u(\Delta\bar u)$ at initial scale $Q_0^2=0.23\,GeV^2$.
        The dotted line corresponds to the fit of Ref. \cite{ansel_kotz} on $h_{1u}$. 
        GRV94 \cite{grv94} parametrization for $q(x)$ and 
        GRSV95 \cite{GRSV95} parametrization for $\Delta q(x)$ are used.
        }
        \label{fig:comparison}
\end{figure}

Recently, for the fist time, the transversity PDF was extracted \cite{ansel_kotz} from
the combined data of HERMES  \cite{hermes}, COMPASS \cite{comp} and 
BELLE \cite{belle}
collaborations. However, because of the rather poor quality of data 
the error band surrounding the fit on $h_1$ is very large (see Fig. \ref{fig:comparison}),
and, besides, 
the authors of Ref. \cite{ansel_kotz} were compelled to apply the large
number of approximations. In particular, the approximation of zero sea
transversity PDF was applied. However, as it was stressed before, 
in the case of proton-proton collisions namely the sea PDFs play the crucial
role. That is why here 
we will apply two versions of evolution model for transversity
instead of the fit from Ref. \cite{ansel_kotz}. First is 
the model where the Soffer inequality is saturated \cite{vogelsang}: 
\be
h_{1q}(x,Q_0^2)=\frac{1}{2}\left[q(x,Q_0^2)+\Delta q(x,Q_0^2)\right],\quad
h_{1\bar q}(x,Q_0^2)=\frac{1}{2}\left[\bar q(x,Q_0^2)+\Delta\bar q(x,Q_0^2)\right]
\ee
at low initial scale ($Q_0^2=0.23GeV^2$), and then $h_{1q}$, $h_{1\bar q}$ are evolved with DGLAP.
Certainly, this model rather gives the upper bound (maximal value) on SSA. In the second version
of evolution model (see \cite{review,anselmino1} and references therein) the valence and sea
transversity PDFs
are assumed to be equal to helicity PDF  $\Delta q$ at the same initial scale (model
scale) and then $h_{1q}$ and $h_{1\bar q}$ are again evolved with  DGLAP to the required $Q^2$ values.
Certainly, this model is much more realistic one because at the model initial
scale a lot of models predict \cite{review} that $h_1=\Delta q$. It is of importance that 
the curve corresponding to this version of evolution model lies just inside 
the error band for the fit of Ref. \cite{ansel_kotz} -- see Fig. \ref{fig:comparison}. 
Thus, this version of evolution model
is consistent with the analysis of Ref. \cite{ansel_kotz}.

\begin{figure}
        \begin{tabular}{cc}
                \includegraphics[height=5cm]{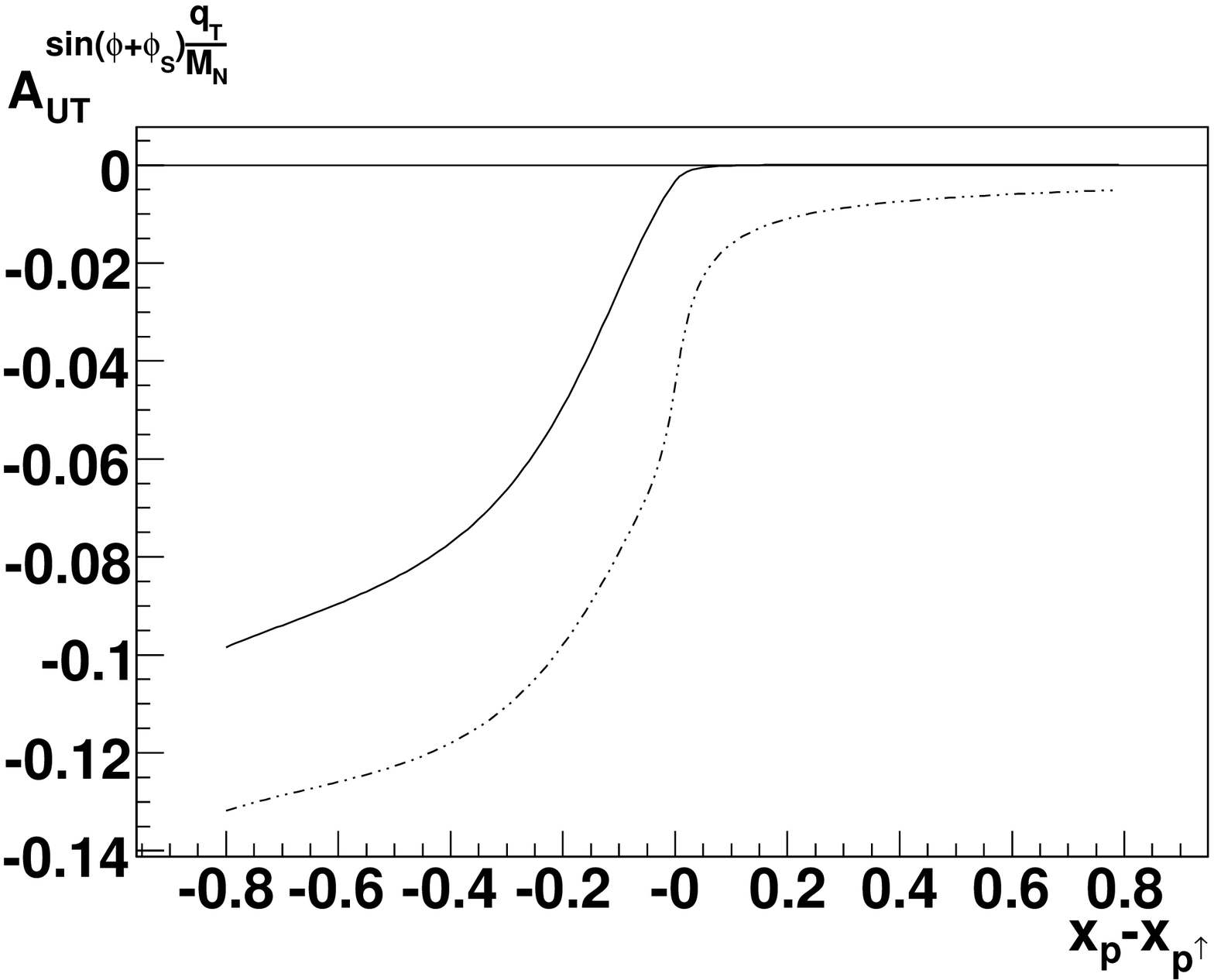} & \includegraphics[height=5cm]{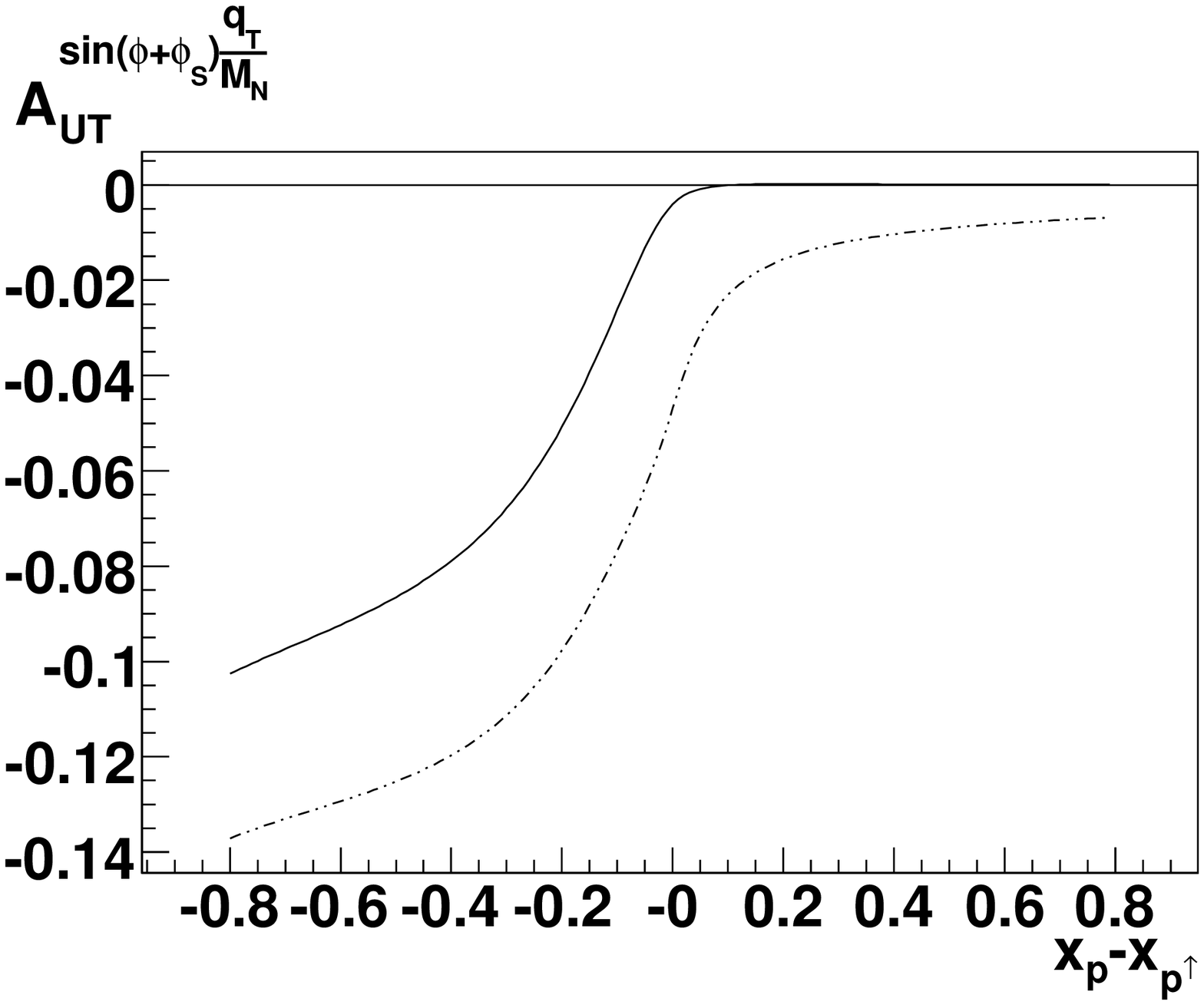} 
        \end{tabular}
 \caption{
Estimation of SSA $A_{UT}^{\sin(\phi+\phi_S)\frac{q_T}{M_N}}\palka_{pp^\uparrow}$ 
        for RHIC, $s=200^2\,GeV^2$,
        with $Q^2=4\,GeV^2$ (left) and $Q^2=20\,GeV^2$ (right). 
The solid and dotted curves correspond to
the two different input ansatzes for $h_{1u}$ which are used in evolution model. 
These are  $h_{1q,\bar q}=\Delta q,\bar q$ and
$h_{1q}=(\Delta q+q)/2$ $h_{1\bar q}=(\Delta \bar q+\bar q)/2$, respectively.
        Here GRV94 \cite{grv94} parametrization for $q(x)$ and 
        GRSV95 \cite{GRSV95} parametrization for $\Delta q(x)$ are used.
}
 \label{fig:bm2}
\end{figure}

\begin{figure}
        \begin{tabular}{cc}
              \includegraphics[height=5cm]{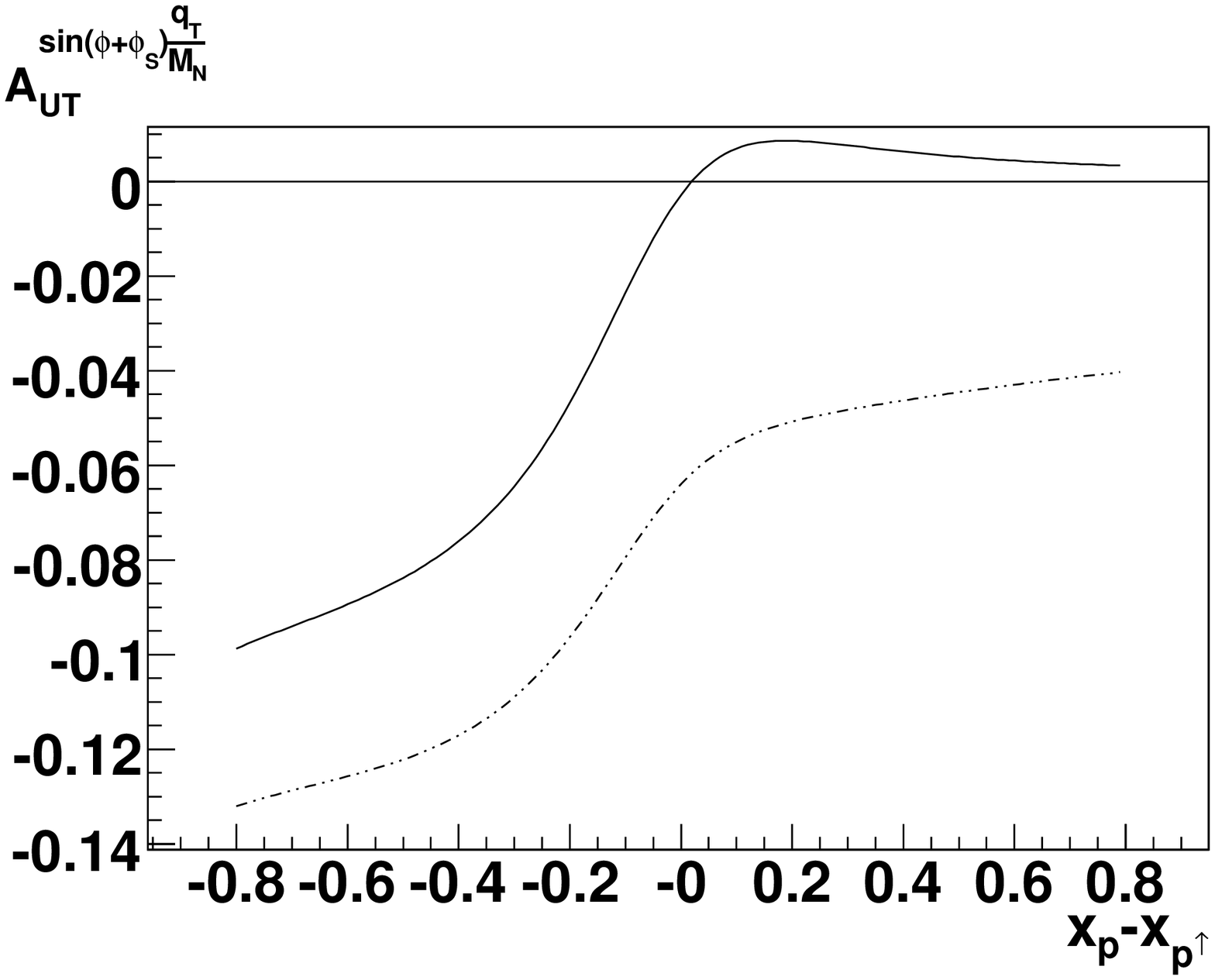} & \includegraphics[height=5cm]{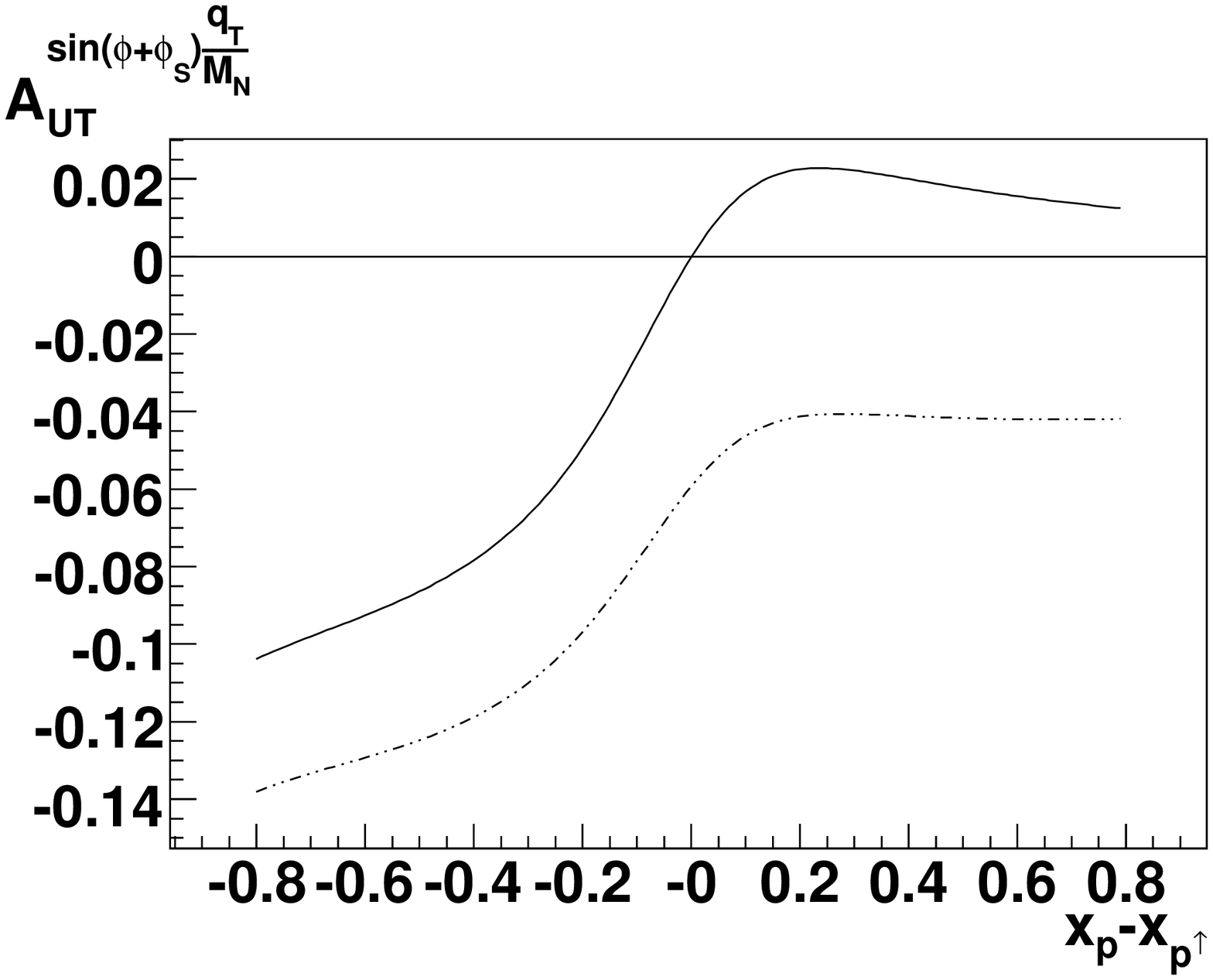} 
        \end{tabular}
        
 \caption{
Estimation of SSA $A_{UT}^{\sin(\phi+\phi_S)\frac{q_T}{M_N}}\palka_{pp^\uparrow}$ 
        for NICA, $s=400\,GeV^2$,
        with $Q^2=4\,GeV^2$ (left) and $Q^2=15\,GeV^2$ (right). 
The solid and dotted curves correspond to
the two different input ansatzes for $h_{1u}$ which are used in evolution model. 
These are  $h_{1q,\bar q}=\Delta q,\bar q$ and
$h_{1q}=(\Delta q+q)/2$ $h_{1\bar q}=(\Delta \bar q+\bar q)/2$, respectively.
        Here GRV94 \cite{grv94} parametrization for $q(x)$ and 
        GRSV95 \cite{GRSV95} parametrization for $\Delta q(x)$ are used.
}
 \label{fig:bm1}
\end{figure}

We present here the estimations of SSA  
$A_{UT}^{\sin(\phi+\phi_S)\frac{q_T}{M_N}}$ 
below and above $J/\psi$ resonance
for the strongly different RHIC (Fig. \ref{fig:bm2}) and NICA (Fig. \ref{fig:bm1}) kinematical
conditions. The respective plots for the COMPASS and J-PARC kinematics are again very 
similar to that for the NICA kinematical range.

Looking at Fig. \ref{fig:bm2} and \ref{fig:bm1} 
one can see that for both RHIC and NICA (as well as for COMPASS
and J-PARC) kinematics 
 the  asymmetry $A_{UT}^{\sin(\phi+\phi_S)\frac{q_T}{M_N}}$ is 
negligible at $x_p > x_{p^\uparrow} $ and
 is quite considerable at $x_p<x_{p^\uparrow}$.  In this second case SSA 
 $A_{UT}^{\sin(\phi+\phi_S)\frac{q_T}{M_N}}$ takes its maximal values (about 5-10\%)
 when $x_p-x_{p^\uparrow}$ takes the large negative values. 
Thus, one can again see the advantage of the symmetric collider mode (RHIC, NICA) where
the cases $x_p<x_{p^\uparrow}$ and $x_p>x_{p^\uparrow}$ 
 do not differ especially. On the contrary, for the fixed target mode these cases
essentially differ because of the acceptance restriction (\ref{restriction}). 
Thus, to obtain
 nonzero $A_{UT}^{\sin(\phi+\phi_S)\frac{q_T}{M_N}}$ with the fixed target, one should either 
 manage to avoid the restriction (\ref{restriction}) dealing with unpolarized 
 beam/polarized target option (forward-backward geometry spectrometer), or, in the case 
of the forward-geometry
spectrometer (acceptance restriction (\ref{restriction}) holds),
 one should deal with the polarized beam/unpolarized target option.
Regretfully, both these options are hardly possible for the running COMPASS experiment.
At the same time, the option with the polarized proton beam is now planned 
at J-PARC facility \cite{sawada}.

In conclusion of this section, 
let us estimate 
how good
are the approximations given by Eqs. (\ref{siv_asympt_01})-(\ref{bm_asympt_1}) 
in the case $x_p\gg x_{p^\uparrow}$, and by Eqs. (\ref{siv_asympt_02})-(\ref{bm_asympt_2})
in the case $x_p\ll x_{p^\uparrow}$.
This is very important for the analysis since these approximations allow us 
to cancel out the extra unknown variables
entering the equations for measured asymmetries.
The respective calculations of $A_{UT}^{\sin(\phi-\phi_S)\frac{q_T}{M_N}}$
are presented\footnote{For better readability of the paper we present here only two tables
for NICA kinematics.
Our calculations show that for all another considered within the paper
kinematical conditions the results are absolutely analogous.} 
by the Table \ref{table:siv_appr}. From this table it is seen that the approximations
(\ref{siv_asympt_01}) and (\ref{siv_asympt_02}) obtained by neglecting the sea PDFs
contributions at high $x$ work very well (column ``B''). At the same time, 
the agreement of the results obtained with the double 
approximations (\ref{siv_asympt_1}) and (\ref{siv_asympt_2}) (column ``C'') 
with the result of column ``A'' is worse (but still not so bad).
This is not surprising since in the applied parametrizations
(\ref{ef1})-(\ref{ef3}) for the Sivers PDFs \cite{efremov_old}, \cite{efremov_new}
the relation $f_{1T}^{\perp(1)u}=-f_{1T}^{\perp(1)d}$ is used, which is argued
within the $1/N_c$ expansion approach \cite{Nc,Nc1}.

On the other hand, in the case of asymmetry $A_{UT}^{\sin(\phi+\phi_S)\frac{q_T}{M_N}}$
both kinds of approximations (\ref{bm_asympt_01}), (\ref{bm_asympt_1}) and
(\ref{bm_asympt_02}), (\ref{bm_asympt_2}) work very well (see Table  \ref{table:bm_appr1}).

\begin{table}[t]
        \caption{ NICA kinematics.
        Values of $A_{UT}^{\sin(\phi-\phi_S)\frac{q_T}{M_N}}$ calculated 
        by using  Eq. (\ref{ssa_siv}) (labeled as A)
        in comparison with two approximations: Eqs. (\ref{siv_asympt_01}), (\ref{siv_asympt_02}), labeled as B, 
        and Eqs. (\ref{siv_asympt_1}),
         (\ref{siv_asympt_2}), labeled as C. For $f_{1T}$ 
        the parametrization  from Ref. \cite{efremov_new} is used.}
\begin{tabular}{cccc}
\hline
\multicolumn{4}{c}{$s=400\,GeV^2$, $Q^2=4\, GeV^2$}\\ 
\hline
$x_p-x_{p^\uparrow}$ & A& B & C\\ 
\hline
-0.4000 & 0.0189 & 0.0184 & 0.0277 \\-0.5000 & 0.0131 & 0.0129 & 0.0190 \\-0.6000 & 0.0087 & 0.0086 & 0.0125 \\-0.7000 & 0.0053 & 0.0053 & 0.0076 \\-0.8000 & 0.0028 & 0.0028 & 0.0040 \\\hline
0.4000 & 0.0514 & 0.0525 & 0.0614 \\0.5000 & 0.0486 & 0.0491 & 0.0556 \\0.6000 & 0.0460 & 0.0462 & 0.0509 \\0.7000 & 0.0437 & 0.0438 & 0.0471 \\0.8000 & 0.0417 & 0.0418 & 0.0439 \\\hline
\end{tabular}
\begin{tabular}{cccc}
\hline
\multicolumn{4}{c}{$s=400\,GeV^2$, $Q^2=15\, GeV^2$}\\ 
\hline
$x_p-x_{p^\uparrow}$ & A& B & C\\ 
\hline
-0.4000 & 0.0178 & 0.0170 & 0.0277 \\-0.5000 & 0.0132 & 0.0129 & 0.0204 \\-0.6000 & 0.0093 & 0.0093 & 0.0142 \\-0.7000 & 0.0061 & 0.0061 & 0.0091 \\-0.8000 & 0.0033 & 0.0033 & 0.0049 \\\hline
0.4000 & 0.0828 & 0.0849 & 0.0984 \\0.5000 & 0.0811 & 0.0820 & 0.0922 \\0.6000 & 0.0788 & 0.0792 & 0.0867 \\0.7000 & 0.0764 & 0.0765 & 0.0818 \\0.8000 & 0.0742 & 0.0742 & 0.0775 \\\hline
\end{tabular}

\label{table:siv_appr}
\end{table}
\begin{table}[t]
        \caption{NICA kinematics.
        Values of $A_{UT}^{\sin(\phi+\phi_S)\frac{q_T}{M_N}}$ calculated by using Eq. (\ref{ssa_bm}) (labeled as A)
        in comparison with two approximations: (\ref{bm_asympt_01}), (\ref{bm_asympt_02}), labeled as B, and
        (\ref{bm_asympt_1}), (\ref{bm_asympt_2}), labeled as C.
        The evolution model for transversity
        with
        $h_{1q(\bar q)}=\Delta q(\Delta\bar q)$ at $Q_0^2=0.23GeV^2$ is used.         
        }
\begin{tabular}{cccc}
\hline
\multicolumn{4}{c}{$s=400\,GeV^2$, $Q^2=4\, GeV^2$}\\ 
\hline
$x_p-x_{p^\uparrow}$ & A& B & C\\ 
\hline
-0.4000 & -0.0761 & -0.0800 & -0.0912 \\-0.5000 & -0.0838 & -0.0856 & -0.0948 \\-0.6000 & -0.0894 & -0.0902 & -0.0975 \\-0.7000 & -0.0940 & -0.0943 & -0.1000 \\-0.8000 & -0.0987 & -0.0988 & -0.1029 \\\hline
0.4000 & 0.0063 & 0.0067 & 0.0068 \\0.5000 & 0.0052 & 0.0054 & 0.0054 \\0.6000 & 0.0044 & 0.0045 & 0.0045 \\0.7000 & 0.0038 & 0.0038 & 0.0038 \\0.8000 & 0.0033 & 0.0033 & 0.0033 \\\hline
\end{tabular}
\begin{tabular}{cccc}
\hline
\multicolumn{4}{c}{$s=400\,GeV^2$, $Q^2=15\, GeV^2$}\\ 
\hline
$x_p-x_{p^\uparrow}$ & A& B & C\\ 
\hline
-0.4000 & -0.0783 & -0.0833 & -0.0951 \\-0.5000 & -0.0864 & -0.0887 & -0.0983 \\-0.6000 & -0.0926 & -0.0936 & -0.1012 \\-0.7000 & -0.0980 & -0.0984 & -0.1041 \\-0.8000 & -0.1038 & -0.1039 & -0.1078 \\\hline
0.4000 & 0.0200 & 0.0216 & 0.0220 \\0.5000 & 0.0176 & 0.0184 & 0.0186 \\0.6000 & 0.0156 & 0.0159 & 0.0160 \\0.7000 & 0.0138 & 0.0139 & 0.0140 \\0.8000 & 0.0123 & 0.0124 & 0.0124 \\\hline
\end{tabular}

\label{table:bm_appr1}
\end{table}
Thus, dealing with the kinematical region $x_p\gg x_{p^\uparrow}$ one  can safely 
use the approximation (\ref{siv_asympt_01}) and, thereby, get the access to the first 
moments of the sea Sivers PDFs. Dealing with the approximation  (\ref{siv_asympt_1})
one should be more careful -- if indeed $f_{1T}^{\perp(1)u}\simeq-f_{1T}^{\perp(1)d}$ 
then double  approximation (\ref{siv_asympt_1}) is suitable only for some rather rough estimations.
At the same time, we have no access to the transversity and Boer-Mulders function in the region
$x_p\gg x_{p^\uparrow}$ since the asymmetry $A_{UT}^{\sin(\phi+\phi_S)\frac{q_T}{M_N}}$
is negligible within this kinematical region. Let us recall that for the fixed target mode,
this kinematical region corresponds to the option with the unpolarized beam and polarized target
(if, certainly, we have in our disposal only the forward-geometry spectrometer -- see Eq. (\ref{restriction})).

Let us now consider another limiting case  $x_p\ll x_{p^\uparrow}$, which in the case
of the fixed target mode and forward-geometry spectrometer corresponds to the option with
the polarized beam and unpolarized target. In this case the situation is absolutely opposite. Here
the asymmetry $A_{UT}^{\sin(\phi+\phi_S)\frac{q_T}{M_N}}$ is quite considerable
(and presumably is measurable as it was discussed above). As it has just shown, in this limit 
we can safely apply even double approximation (\ref{bm_asympt_2}). This give us the interesting 
possibility (see Section 2) to extract the ratio $h_{1u}/h_{1u}^{\perp(1)}$ directly, without application of 
fitting procedure with a set of assumptions on extra unknown variables.

\section{SSA in $pD$ and $DD$ collisions }
As usual, the inclusion of the deutron beam/target can allow us 
to find PDFs of $u$ and $d$ quark, in separation.

Applying $SU_f(2)$ symmetry to the results 
of Section 3
one immediately gets the respective results on SSA for Drell-Yan 
processes in $ pD$ and $DD$ collisions.
For SSA giving an access to the Sivers PDF,
in the limiting case (\ref{limit1}) one gets instead of Eq. (\ref{siv_asympt_01}) the 
equations
\be
\label{siv_polPunpolD_1}
A_{UT}^{\sin(\phi-\phi_S)\frac{q_T}{M_N}}(x_D\gg x_{p^\uparrow})\palka_{Dp^\uparrow \rightarrow l^+l^-X}  \simeq  
2 \frac{
4 \bar f_{1T}^{\perp(1)u}(x_{p^\uparrow}) +  \bar f_{1T}^{\perp(1)d}(x_{p^\uparrow}) }
{4\bar f_{1u}(x_{p^\uparrow})+\bar f_{1d}(x_{p^\uparrow})},
\ee 
and
\be
\label{siv_unpolPpolD_1}
A_{UT}^{\sin(\phi-\phi_S)\frac{q_T}{M_N}}(x_p\gg x_{D^\uparrow})\palka_{pD^\uparrow \rightarrow l^+l^-X}  
&=&A_{UT}^{\sin(\phi-\phi_S)\frac{q_T}{M_N}}(x_D\gg x_{D^\uparrow})\palka_{DD^\uparrow \rightarrow l^+l^-X}\nonumber \\
&\simeq&  
2 \frac{
\bar f_{1T}^{\perp(1)u}(x_{D^\uparrow}) +  \bar f_{1T}^{\perp(1)d}(x_{D^\uparrow}) }
{\bar f_{1u}(x_{D^\uparrow})+\bar f_{1d}(x_{D^\uparrow})},
\ee    
while in the limiting case (\ref{limit2}) one obtains

\be
\label{siv_polPunpolD_2}
A_{UT}^{\sin(\phi-\phi_S)\frac{q_T}{M_N}}(x_D\ll x_{p^\uparrow})\palka_{Dp^\uparrow \rightarrow l^+l^-X}  \simeq  
2 \frac{
4  f_{1T}^{\perp(1)u}(x_{p^\uparrow}) +  f_{1T}^{\perp(1)d}(x_{p^\uparrow}) }
{4 f_{1u}(x_{p^\uparrow})+ f_{1d}(x_{p^\uparrow})},
\ee 
and
\be
\label{siv_unpolPpolD_2}
A_{UT}^{\sin(\phi-\phi_S)\frac{q_T}{M_N}}(x_p\ll x_{D^\uparrow})\palka_{pD^\uparrow \rightarrow l^+l^-X}  
&=&A_{UT}^{\sin(\phi-\phi_S)\frac{q_T}{M_N}}(x_D\ll x_{D^\uparrow})\palka_{DD^\uparrow \rightarrow l^+l^-X}\nonumber \\
&\simeq&  
2 \frac{
f_{1T}^{\perp(1)u}(x_{D^\uparrow}) +   f_{1T}^{\perp(1)d}(x_{D^\uparrow}) }
{ f_{1u}(x_{D^\uparrow})+ f_{1d}(x_{D^\uparrow})}.
\ee    
At the same time, SSA giving an access to transversity and Boer-Mulders PDFs
look as
\be
\label{bm_polPunpolD_1}
A_{UT}^{\sin(\phi+\phi_S)\frac{q_T}{M_N}}(x_D\gg x_{p^\uparrow})\palka_{Dp^\uparrow \rightarrow l^+l^-X}  \simeq 
- \frac{[h_{1u}^{\perp(1)}(x_D)+ h_{1d}^{\perp(1)}(x_D)]
[4\bar h_{1u}(x_{p^\uparrow})+\bar h_{1d}(x_{p^\uparrow})]}
{[f_{1u}(x_D)+f_{1d}(x_D)][ 4\bar f_{1u}(x_{p^\uparrow}) + \bar f_{1d}(x_{p^\uparrow})]},
\ee
\be
\label{bm_unpolPpolD_1}
A_{UT}^{\sin(\phi+\phi_S)\frac{q_T}{M_N}}(x_p\gg x_{D^\uparrow})\palka_{pD^\uparrow \rightarrow l^+l^-X}  \simeq 
- \frac{[4h_{1u}^{\perp(1)}(x_p)+ h_{1d}^{\perp(1)}(x_p)]
[\bar h_{1u}(x_{D^\uparrow})+\bar h_{1d}(x_{D^\uparrow})]}
{[4f_{1u}(x_p)+f_{1d}(x_p)][\bar f_{1u}(x_{D^\uparrow}) + \bar f_{1d}(x_{D^\uparrow})]},
\ee
and
\be
\label{bm_unpolDpolD_1}
A_{UT}^{\sin(\phi+\phi_S)\frac{q_T}{M_N}}(x_D\gg x_{D^\uparrow})\palka_{DD^\uparrow \rightarrow l^+l^-X}  \simeq 
- \frac{[ h_{1u}^{\perp(1)}(x_D)+ h_{1d}^{\perp(1)}(x_D)]
[\bar h_{1u}(x_{D^\uparrow})+\bar h_{1d}(x_{D^\uparrow})]}
{[ f_{1u}(x_D)+f_{1d}(x_D)][\bar f_{1u}(x_{D^\uparrow}) + \bar f_{1d}(x_{D^\uparrow})]}
\ee
in the limiting case (\ref{limit1}), while
\be
\label{bm_polPunpolD_2}
A_{UT}^{\sin(\phi+\phi_S)\frac{q_T}{M_N}}(x_D\ll x_{p^\uparrow})\palka_{Dp^\uparrow \rightarrow l^+l^-X}  \simeq 
- \frac{[\bar h_{1u}^{\perp(1)}(x_D)+ \bar h_{1d}^{\perp(1)}(x_D)]
[4 h_{1u}(x_{p^\uparrow})+ h_{1d}(x_{p^\uparrow})]}
{[\bar f_{1u}(x_D)+\bar f_{1d}(x_D)][ 4 f_{1u}(x_{p^\uparrow}) + f_{1d}(x_{p^\uparrow})]},
\ee
\be
\label{bm_unpolPpolD_2}
A_{UT}^{\sin(\phi+\phi_S)\frac{q_T}{M_N}}(x_p\ll x_{D^\uparrow})\palka_{pD^\uparrow \rightarrow l^+l^-X}  \simeq 
- \frac{[4\bar h_{1u}^{\perp(1)}(x_p)+ \bar h_{1d}^{\perp(1)}(x_p)]
[ h_{1u}(x_{D^\uparrow})+ h_{1d}(x_{D^\uparrow})]}
{[4\bar f_{1u}(x_p)+\bar f_{1d}(x_p)][ f_{1u}(x_{D^\uparrow}) + f_{1d}(x_{D^\uparrow})]},
\ee
and
\be
\label{bm_unpolDpolD_2}
A_{UT}^{\sin(\phi+\phi_S)\frac{q_T}{M_N}}(x_D\ll x_{D^\uparrow})\palka_{DD^\uparrow \rightarrow l^+l^-X}  \simeq 
- \frac{[ \bar h_{1u}^{\perp(1)}(x_D)+ \bar h_{1d}^{\perp(1)}(x_D)]
[h_{1u}(x_{D^\uparrow})+ h_{1d}(x_{D^\uparrow})]}
{[ \bar f_{1u}(x_D)+\bar f_{1d}(x_D)][f_{1u}(x_{D^\uparrow}) +  f_{1d}(x_{D^\uparrow})]}
\ee
in the limiting case (\ref{limit2}).

As it was mentioned above (see Section 3) 
there exist the strong theoretical arguments \cite{Nc}, \cite{Nc1} 
based on $1/N_{c}$ expansion
that the sum of the $u$ and $d$ quark Sivers first moments
$f_{1T}^{\perp(1)u}$  and $f_{1T}^{\perp(1)d}$  
is also very small quantity\footnote{That is why the equality
$f_{1T}^{\perp(1)u}+ f_{1T}^{\perp(1)d}=0$ in the parametrizations
(\ref{ef1})-(\ref{ef3}) is applied. Together with the assumption 
(\ref{seamodel1}) it denotes that the equality 
$\bar f_{1T}^{\perp(1)u}+ \bar f_{1T}^{\perp(1)d}=0$
for the sea PDFs
is also holds for these parametrizations.}. Besides, 
the QCD evolution predicts small values of the sea transversity
distributions {\it even at small} $x$ values \cite{anselmino1}.
Thus, in the case of polarized deutron in initial state, almost all respective SSA
(see (\ref{siv_unpolPpolD_1}), (\ref{siv_unpolPpolD_2}), 
(\ref{bm_unpolPpolD_1}), (\ref{bm_unpolDpolD_1}))
presumably should be very small quantities (and our calculations confirm it), compatible with zero
within the errors (certainly, it should be carefully checked by the respective measurements
at RHIC, NICA, COMPASS and J-PARC).  
The only SSA which could
take considerable value are SSA containing the sum 
$h_{1u}(x_{D^\uparrow})+h_{1d}(x_{D^\uparrow})$
(see Eqs. (\ref{bm_unpolPpolD_2}),  (\ref{bm_unpolDpolD_2})). The point is that the analysis
\cite{ansel_kotz} of the COMPASS data \cite{comp}
obtained on the deutron target produced the possibility of nonzero sum
$h_{1u}+h_{1d}$. In accordance with this analysis
$h_{1u}$ and $h_{1d}$ are of opposite sign but differ\footnote{Evolution
model, which is consistent with the analysis of Ref. \cite{ansel_kotz}
(see discussion around Fig. \ref{fig:comparison}), also predicts that
$h_{1u}\ne h_{1d}$.} in their absolute
values (see Fig. 7 in Ref. \cite{ansel_kotz}). However, the uncertainties on 
$h_{1u}$ and $h_{1d}$ are too large (see the error bands in Fig. 7) to realize
is the quantity $h_{1u}+h_{1d}$ zero or not. Thus, the respective measurements
of SSA in DY processes with polarized deutron could shed the light on this problem.

\begin{figure}
        \begin{tabular}{cc}
                \includegraphics[height=5cm]{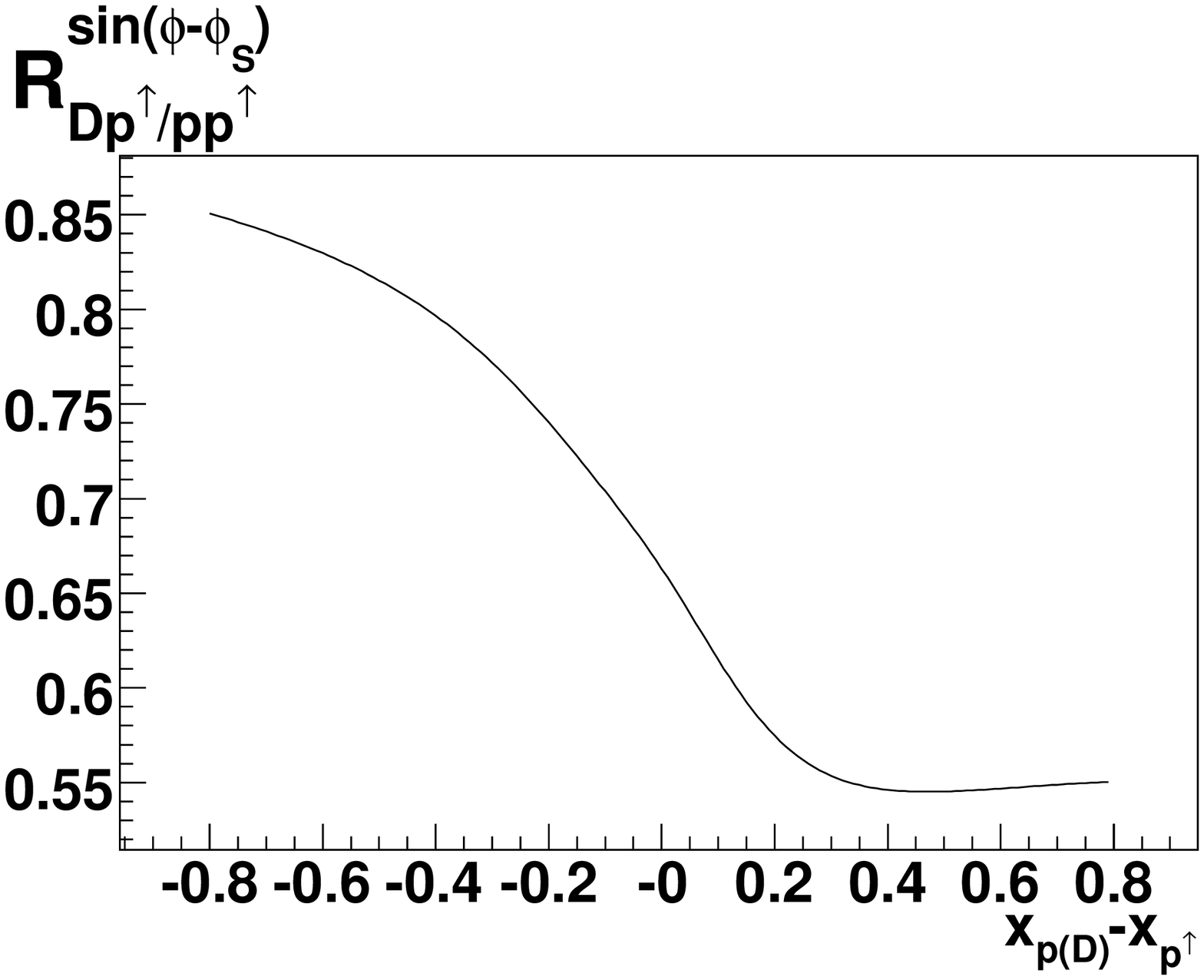} & \includegraphics[height=5cm]{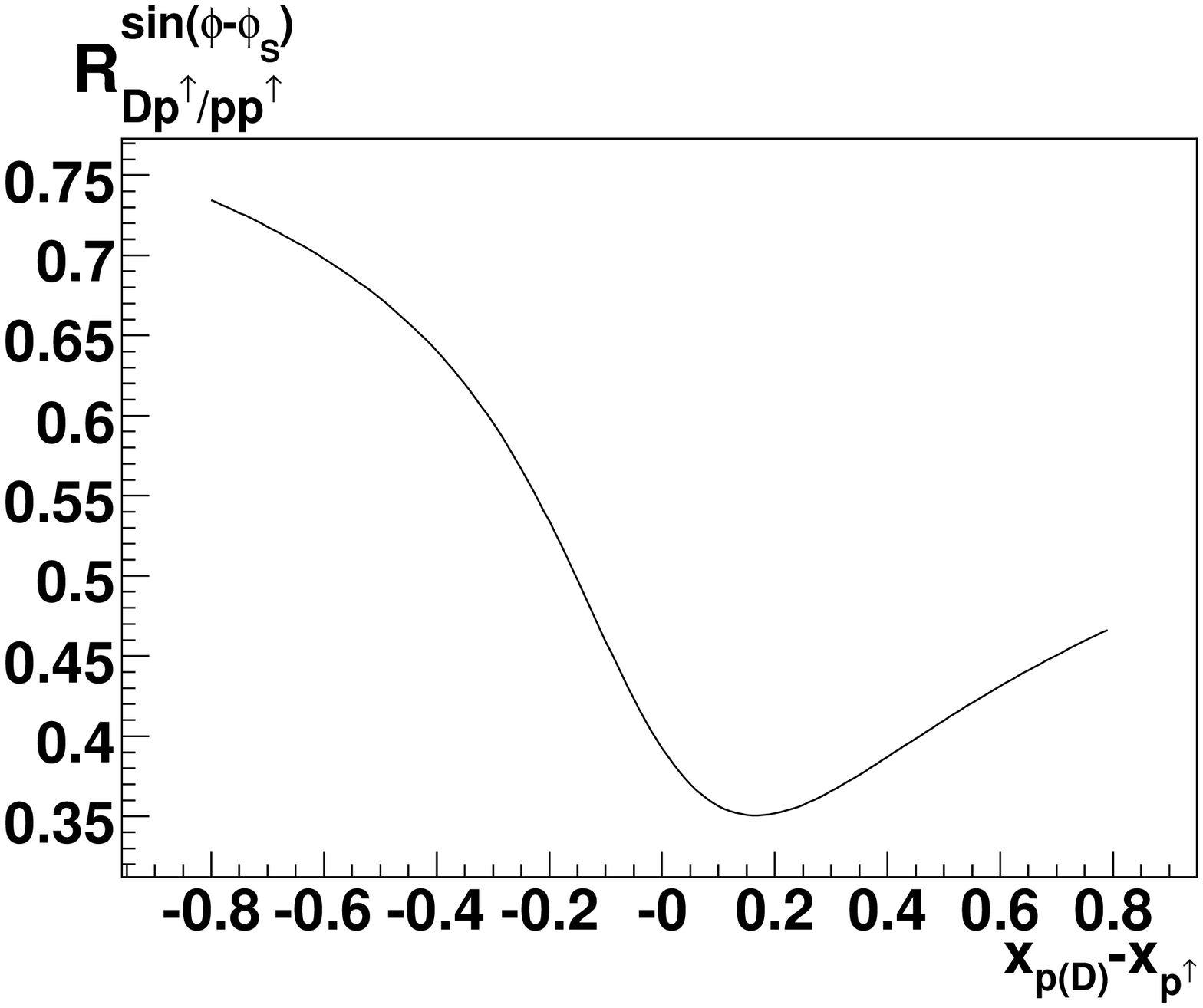}         

        \end{tabular}
 \caption{
Estimation of ratio $
R=A_{UT}^{\sin(\phi-\phi_S)\frac{q_T}{M_N}}{\Bigl |}_{Dp^\uparrow}/A_{UT}^{\sin(\phi-\phi_S)\frac{q_T}{M_N}}{\Bigl |}_{pp^\uparrow}
$ for NICA kinematics with $Q^2=4\,GeV^2$ (left) and   $Q^2=15\,GeV^2$ (right).
}
 \label{fig:dp_sivers}
\end{figure}

\begin{figure}
        \begin{tabular}{cc}
                \includegraphics[height=5cm]{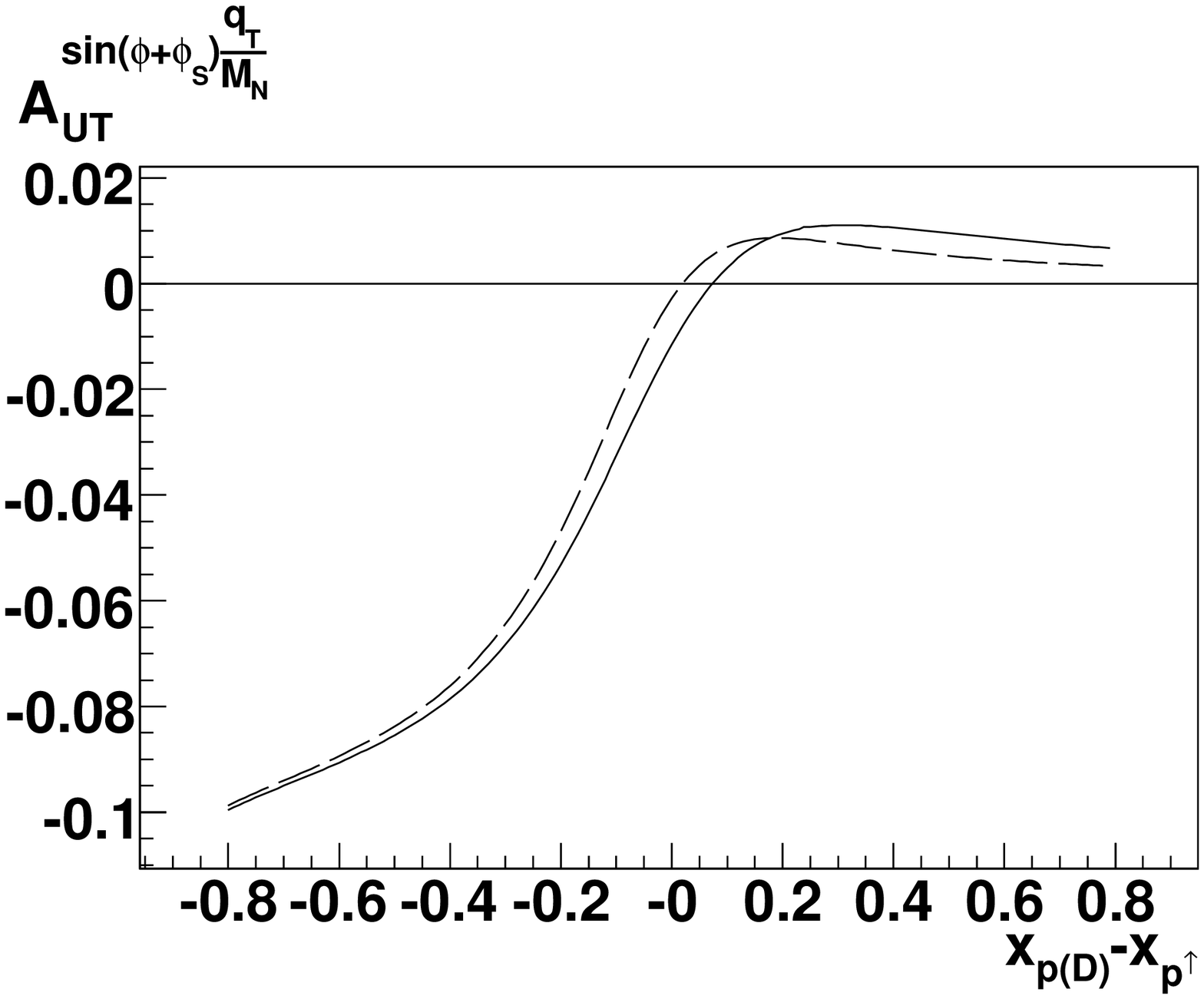} & \includegraphics[height=5cm]{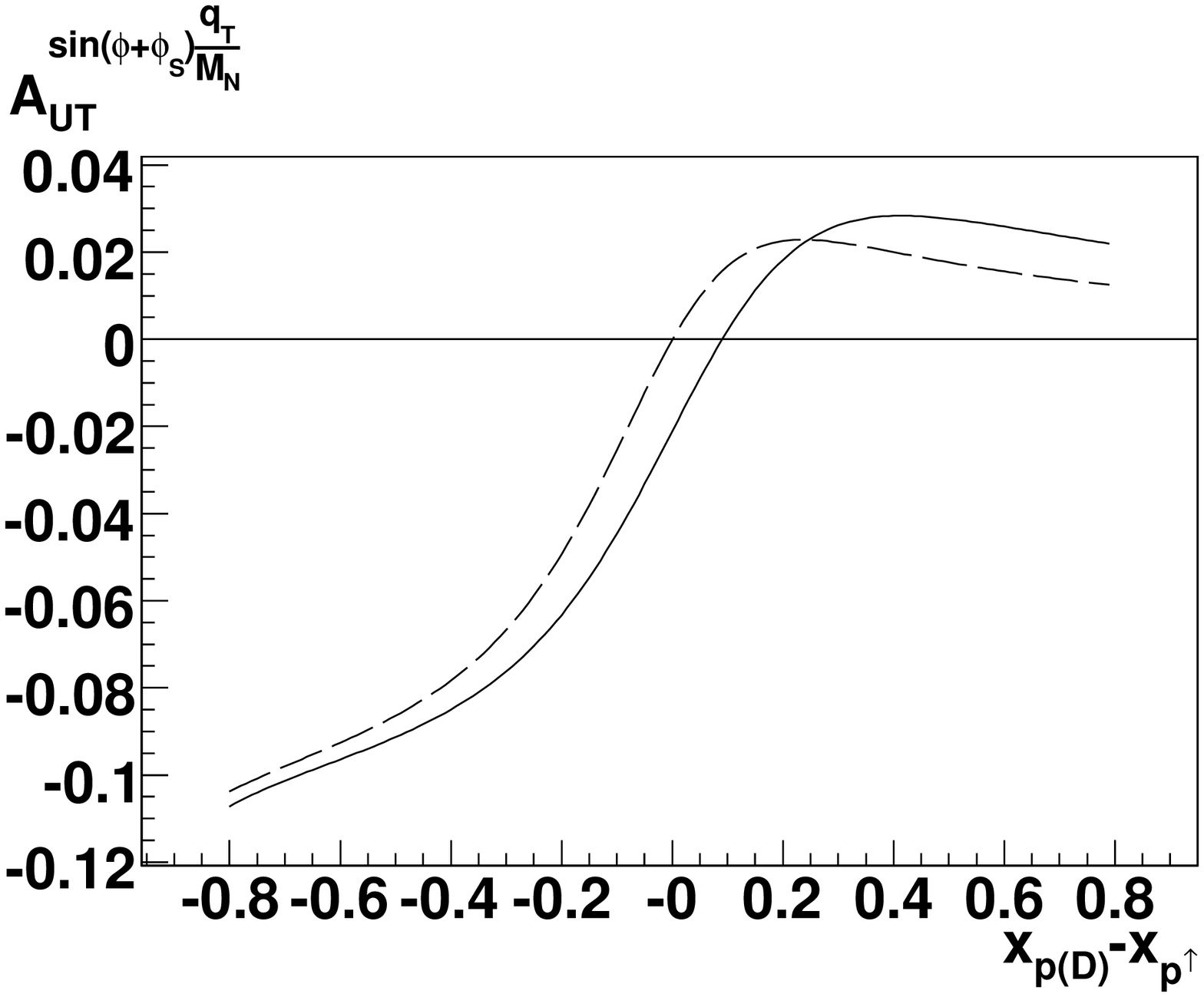}

        \end{tabular}
 \caption{
Estimation of SSA $A_{UT}^{\sin(\phi+\phi_S)\frac{q_T}{M_N}}$ 
        for NICA with two different options, 
        $pp^\uparrow$ (dashed line) and $Dp^\uparrow$ (solid line) collisions,
        with $Q^2=4\,GeV^2$ (left) and $Q^2=15\,GeV^2$ (right). 
        Here evolution model for $h_{1q}$ is used with the input ansatz
  $h_{1q,\bar q}=\Delta q,\bar q$.
        GRV94 \cite{grv94} for $q$ and 
        GRSV95 \cite{GRSV95} parametrizations for $\Delta q$ are used.
}
 \label{fig:dp_bm}
\end{figure}

On the contrary to the case of polarized deutron, in the case of polarized proton
all SSA could take the considerable values. Our calculations
show that they are of the same order of magnitude as the respective
SSA in the case of $pp^\uparrow$ collisions -- see Figs \ref{fig:dp_sivers}, \ref{fig:dp_bm}. 
Notice that for brevity we present the relations between SSA for $pp^\uparrow$
and $Dp^\uparrow$ collisions 
only for NICA
since they are very similar (practically the same) to that for RHIC, COMPASS and J-PARC kinematics.
In Fig. \ref{fig:dp_sivers}  we present the ratio 
$
R=A_{UT}^{\sin(\phi-\phi_S)\frac{q_T}{M_N}}{\Bigl |}_{Dp^\uparrow}/A_{UT}^{\sin(\phi-\phi_S)\frac{q_T}{M_N}}{\Bigl |}_{pp^\uparrow}
$. This ratio (which is the same for all three used parametrizations
for Sivers function) changes from $0.4$ to $0.8$. In Fig. \ref{fig:dp_bm} we present 
the asymmetry $A_{UT}^{\sin(\phi+\phi_S)\frac{q_T}{M_N}}$
for $Dp^\uparrow$ and $pp^\uparrow$ collisions and it is seen that the respective 
curves almost merge. Thus, one can conclude that in the
case of $Dp^\uparrow$ collisions both, weighted with
$\sin(\phi-\phi_S)$ and weighted with $\sin(\phi+\phi_S)$, SSA are presumably measurable
in the same $x$ regions as the respective SSA in the case of $pp^\uparrow$
collisions.

\section{Estimations on the SSA feasibility with the new generator of polarized
DY events}
Generator of polarized Drell-Yan events is necessary for {\it first}, 
estimation of the SSA feasibility on the preliminary (theoretical) stage 
(without details of experimental setup) -- just this paper. {\it  Second,} 
as an input for detector simulation software (for example, GEANT \cite{geant} based code) on both
planning of experimental setup and the data analysis stages.
Until recently there was no in the free access any generator of Drell-Yan events except for the only
PYTHIA generator \cite{pythia}. However, regretfully, in PYTHIA there are only unpolarized Drell-Yan
processes and, besides, they are implemented in PYTHIA without correct $q_T$ and $\cos2\phi$ dependence,
which is absolutely necessary to study Boer-Mulders effect. Recently did appear the first generator
of polarized  DY events \cite{bianconi}, \cite{bianconi1} where $q_T$ and angle dependencies are properly taken into account.

To simplify implementation of the new possibilities in program and to effectively
control all calculations we wrote the new generator of polarized DY events (the details will be published
elsewhere). The scheme of generator is quite simple and very similar to the event generator GMC\_TRANS \cite{gmc_trans} 
which was successfully used by HERMES collaboration for simulation of 
the Sivers effect in semi-inclusive DIS processes \cite{hermes_papers}. Briefly, the scheme of DY event generation
look as follows. First, the generator performs the choice of flavor $q$ of annihilating $q\bar q$ pair and choice does the
given hadron (for example, polarized) contains annihilating quark or, alternatively, antiquark of given (chosen) flavor.  
It is done in accordance with the total unpolarized DY cross-sections for each flavor and
each alternative choice for given hadron in initial state (annihilating quark or antiquark inside).
Then, the variables $x_F$ and $Q^2$ are selected according to the part of unpolarized cross-section (see, for example,
Eq. (1) in Ref. \cite{approach}) which does not contain the angle dependencies. At the next step the polar angle 
$\theta$ is  selected from $\sin\theta(1+\cos^2\theta)$ distribution in that cross-section.  
Then, the Gaussian model for $f_{1q}(x,k_T)$ is applied and after that the transverse momentum of lepton pair $q_T$ 
is selected from exponential distribution $\exp(-q_T^2/2)/2\pi$. At the next step  
$\phi$ and $\phi_S$ angles are selected in accordance with $\cos 2\phi$, $\sin (\phi-\phi_S)$ and $\sin(\phi+\phi_S)$
dependencies of the single-polarized DY cross-section (see, for example, Eq. (2) in Ref. \cite{approach}). 
The $k_T$ dependencies of Boer-Mulders $h_{1q}^\perp(x,k_T)$ and Sivers $f_{1T}^q(x,k_T)$ PDFs are 
fixed by the Boer model \cite{bmodel} and Gaussian ansatz \cite{efremov_old, efremov_new}, respectively. 
At this stage of $\phi$ and $\phi_S$ selection  
$x_F$, $Q^2$, $\theta$ and $q_T$ variables are already fixed\footnote{Certainly, 
one can select all variables simultaneously. However, such scheme essentially decreases the rate of events
generation.} that essentially increases
the rate of $\phi$ and $\phi_S$ selection.   
All variables are generated using the standard von Neumann acception-rejection technique (see, for example \cite{pythia}).

Let us stress once again that the generator elaborated in  \cite{bianconi,bianconi1} is the first
generator of polarized DY events and in many respects it helped us to construct our generator. 
Now we briefly consider the advantages of the new generator.  
 As it was discussed above, one of the main requirement to any 
generator of polarized DY events is to properly include the nontrivial $q_T$ dependence
of DY cross-sections. This is of especial importance for the $q_T$ weighted objects we deal 
with in the paper.
In the earlier generator \cite{bianconi, bianconi1}
the approximations are applied for the convolution calculations in the $h_{1q}^{\perp}$ containing parts of 
DY cross-section just as it was done in the original
paper \cite{bmodel} (see the discussion around Eqs. (47) and (53) in Ref. \cite{bmodel}). 
However, the direct calculations show that implementation of these approximation in the generator 
leads to essential distortion of the $q_T$ weighted objects,  
such as SSA $A_{UT}^{\sin(\phi+\phi_S)\frac{q_T}{M_N}}$: the values of SSA obtained
from simulated data significantly differ from the respective values 
calculated directly from the input parametrizations/models for $f_{1q}$, $h_{1q}^{\perp(1)}$ and $h_{1q}$. 
That is why we avoid the such kind of
approximations. Instead, we calculate the convolutions numerically for the large discrete set of $q_T$ values
and then we perform the standard spline interpolation procedure to reconstruct the calculated convolution
as the continuous function of $q_T$. As a result (see Figs. 7-10  below), 
the values of SSA  $A_{UT}^{\sin(\phi+\phi_S)\frac{q_T}{M_N}}$ reconstructed from simulated data
are in a good agreement with the respective values calculated directly from the parametrizations/models entering
the generator as an input.

Another advantage of the new generator is that it can be much more easy
combined\footnote{This work now in progress.} with the PYTHIA generator (where
almost all possible processes are included)  which is necessary for the
analysis of background processes that can produce
false DY events (misidentification of lepton pair). 
The point is that, on the contrary to generator \cite{bianconi, bianconi1}, in our generator, just as in PYTHIA, the DY processes are generated for
each flavor in separation. 
Besides, the constructed generator have some technical advantages: due to the chosen
generation scheme the rate of event generation is much higher than for generator 
\cite{bianconi, bianconi1}, where all kinematical variables are thrown simultaneously;
the search for cross-sections maximums for the von Neumann algorithm  is performed automatically
as well as the calculation of the total cross-section at the end of each generation run.

Having in our disposal the generator of polarized DY events, we can now estimate the feasibility
of SSA calculated before. Certainly, these are very preliminary estimations on the first (theoretical
level). To perform the comprehensive feasibility  estimations one needs to take into account 
the all peculiarities of the concrete experimental setup.

We prepared two samples with applied statistics 100K and 50K 
of pure Drell-Yan events for each of two  
$Q^2$ ranges:  $2<Q^2<8.5\, GeV^2$ and $Q^2>11\, GeV^2$.  Cut $2<Q^2<8.5\,GeV^2$
is applied to avoid misidentification of lepton pairs due to 
numerous background processes (combinatorial background from Dalitz-decays
and gamma conversions, etc -- see, for instance, section F.4.2 in
Ref. \cite{pax_proposal}) below $Q^2=2\, GeV^2$  and to exclude lepton pairs coming from $J/\psi$ region. 
Cut $Q^2>11\,GeV^2$ is also applied to avoid the lepton pairs coming from $J/\psi$ region.

\begin{figure}
        \begin{tabular}{cc}
\includegraphics[height=5cm]{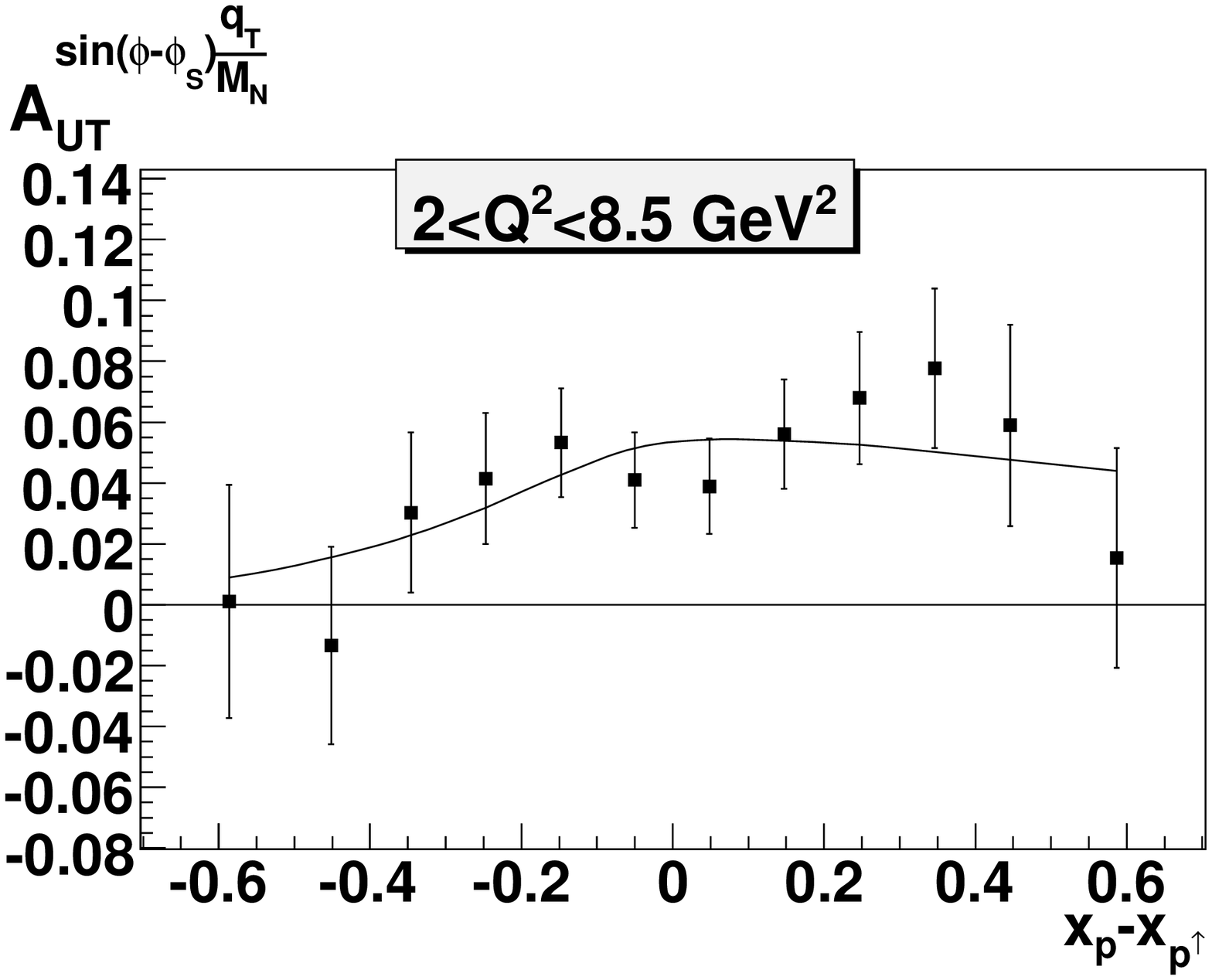} & 
\includegraphics[height=5cm]{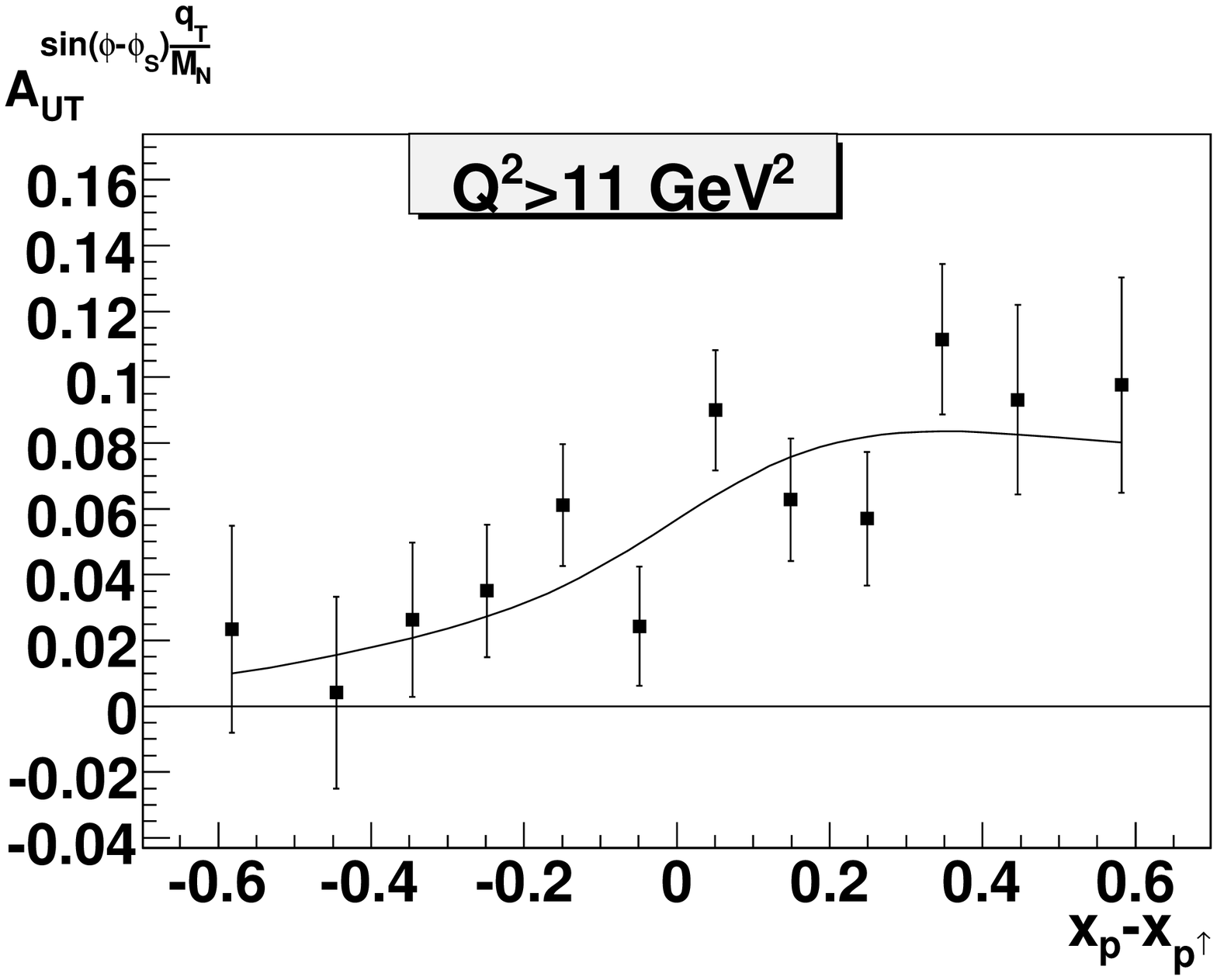}\\
\includegraphics[height=5cm]{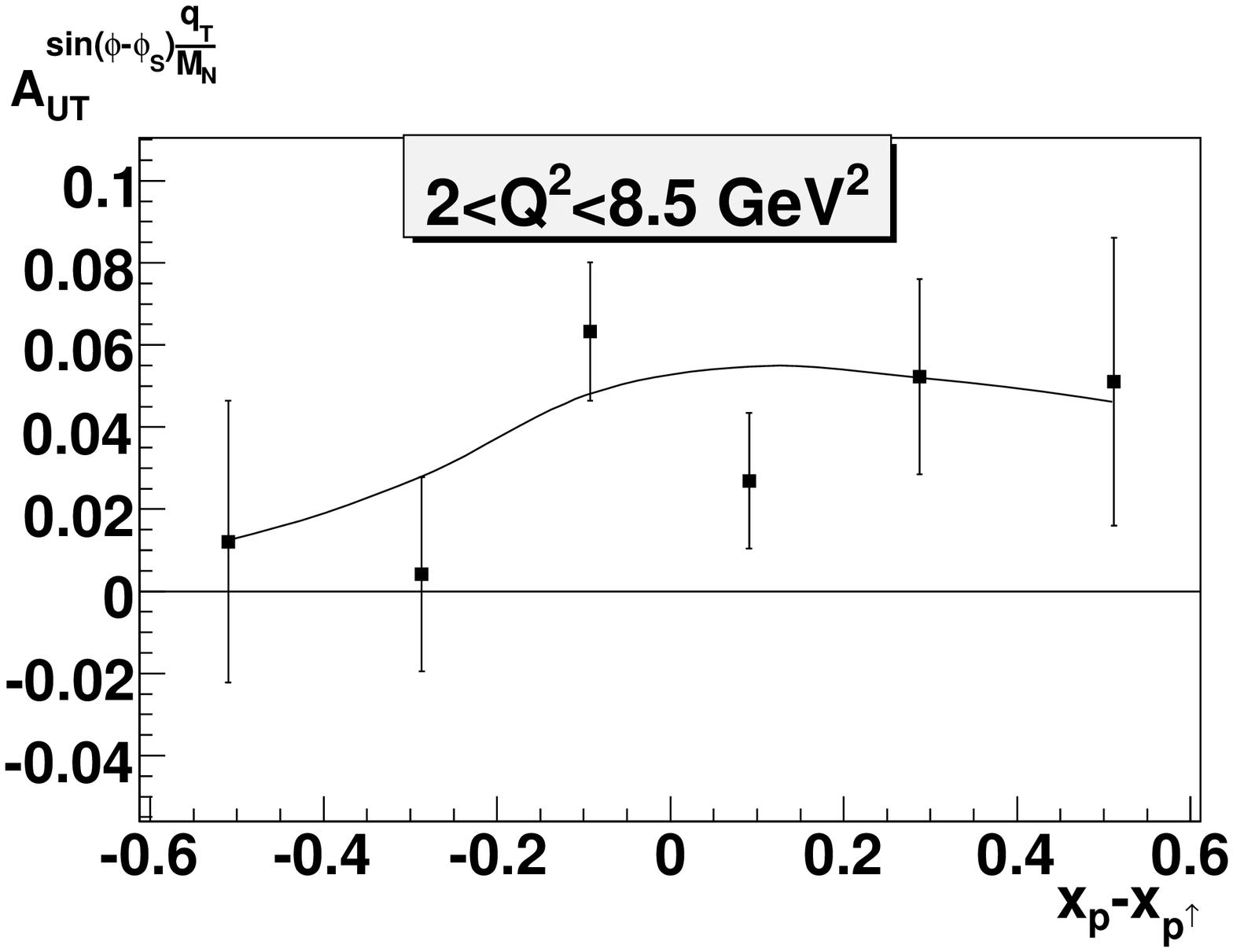} & 
\includegraphics[height=5cm]{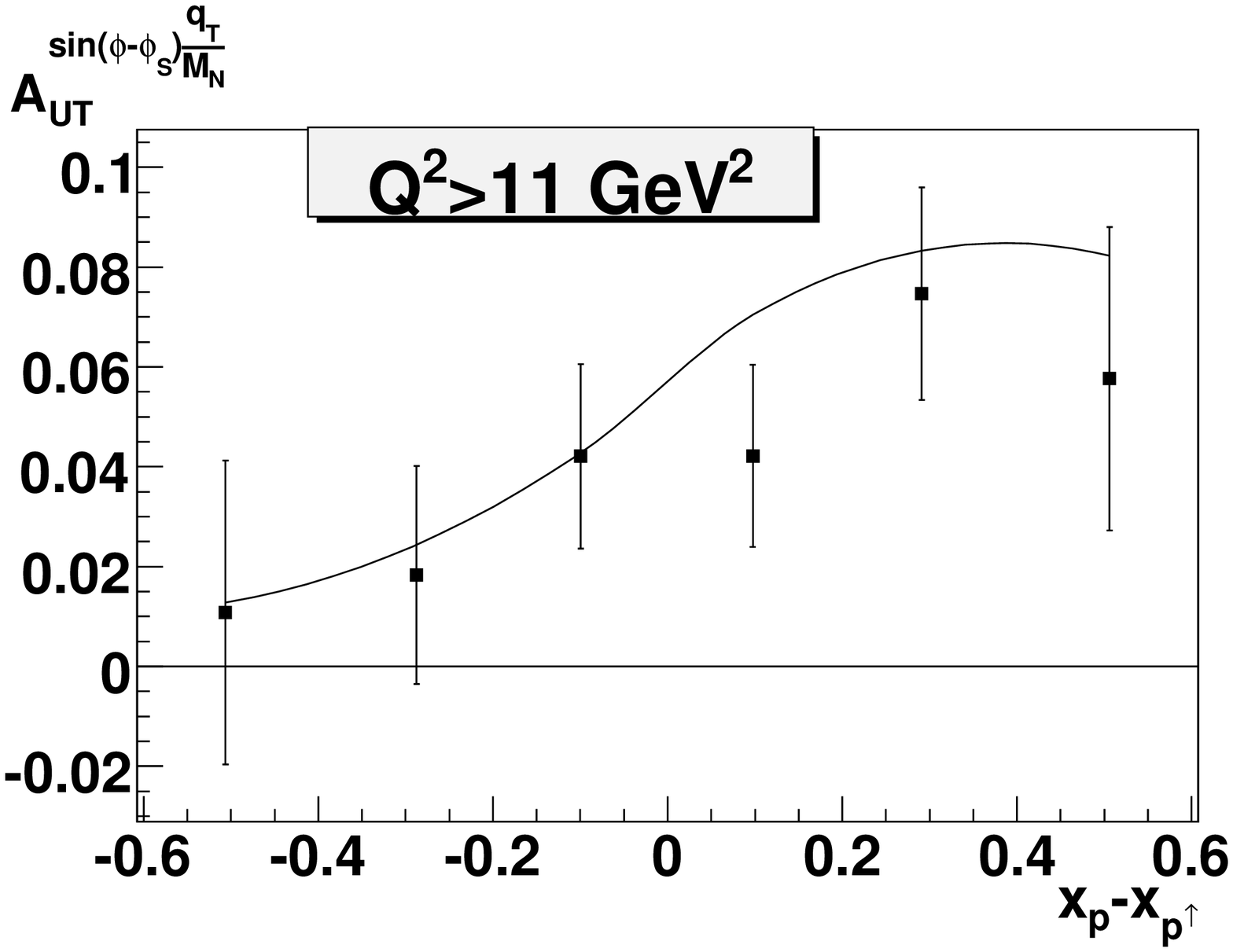}

        \end{tabular}
        \caption{Estimation of asymmetries $A_{UT}^{\sin(\phi-\phi_S)\frac{q_T}{M_N}}\palka_{pp^\uparrow}$  for NICA, $s=400GeV^2$. 
        Here fit   from  Ref. \cite{efremov_new} is used.
The points with error bars are obtained by using simulations with event generator at the applied
statistics 100K pure Drell-Yan events. 
$\langle Q^2\rangle\simeq3.5\,GeV^2$ and $\langle Q^2\rangle=15\,GeV^2$ for the left and right plots, respectively. 
}
\label{fig:simulations_sivers_100k}
\end{figure}

As before, we did not present here the estimation on feasibility of  
$A_{UT}^{\sin(\phi-\phi_S)\frac{q_T}{M_N}}\palka_{pp^\uparrow}$ for RHIC kinematics because it was in detail done
in Ref. \cite{efremov_new}. So, for this SSA we again (see Section 3) present here the results only for the
NICA center of mass energy $20\,GeV$  having in mind that the results on SSA for COMPASS and J-PARC 
(in comparison with RHIC rather close in $\sqrt{s}$ to NICA) occur quite similar (see Section 3) to the 
respective results for NICA. The results are presented in Fig. \ref{fig:simulations_sivers_100k}. 
For the simulations with the developed generator  we use the latest parametrization, fit III (solid line
in Fig. \ref{fig:simulations_sivers_100k}), from the set (\ref{ef1})-(\ref{ef3}).
Looking at Fig. \ref{fig:simulations_sivers_100k} one can see that even at relatively low 
applied statistics 50K pure Drell-Yan events (bottom part of   Fig. \ref{fig:simulations_sivers_100k})
 there are three presumably measurable points for $A_{UT}^{\sin(\phi-\phi_S)\frac{q_T}{M_N}}$  
in the kinematical region $x_p-x_{p^\uparrow}>0$, where this SSA is about 4-6\%. 
Moreover,  at 
applied statistics 100K pure Drell-Yan events one can hope to reconstruct the functional form 
of SSA  $A_{UT}^{\sin(\phi-\phi_S)\frac{q_T}{M_N}}$ (see top part of Fig. \ref{fig:simulations_sivers_100k})
in the kinematical region $x_p>x_{p^\uparrow}$. Since kinematical region $x_p>x_{p^\uparrow}$ corresponds
to $x_F>0$ for the fixed target mode with unpolarized beam and polarized target available to COMPASS
and J-PARC, one can conclude that all four RHIC, NICA, COMPASS and J-PARC
facilities can provide us the access to the Sivers PDF (see Section 3). 
In the region $x_p<x_{p^\uparrow}$ available to RHIC, NICA and (presumably) J-PARC,  
SSA  $A_{UT}^{\sin(\phi-\phi_S)\frac{q_T}{M_N}}$ is smaller (less than 4\%), but still 
visible within the errors (even at applied statistics 50K events one can see at least
one measurable point).

Let us now consider feasibility of the Sivers PDFs in the case of the deutron in initial state.
Looking at Figs. \ref{fig:dp_sivers} 
and \ref{fig:simulations_sivers_100k} one can conclude 
that the only  SSA $A_{UT}^{\sin(\phi-\phi_S)\frac{q_T}{M_N}}\palka_{Dp^\uparrow}$ 
which could take considerable values (see Section 4) is hardly measurable at applied
statistics 50K pure Drell-Yan events. At the same time, in the case of 100K events 
this asymmetry becomes presumably measurable.

\begin{figure}
        \begin{tabular}{cc}
\includegraphics[height=5cm]{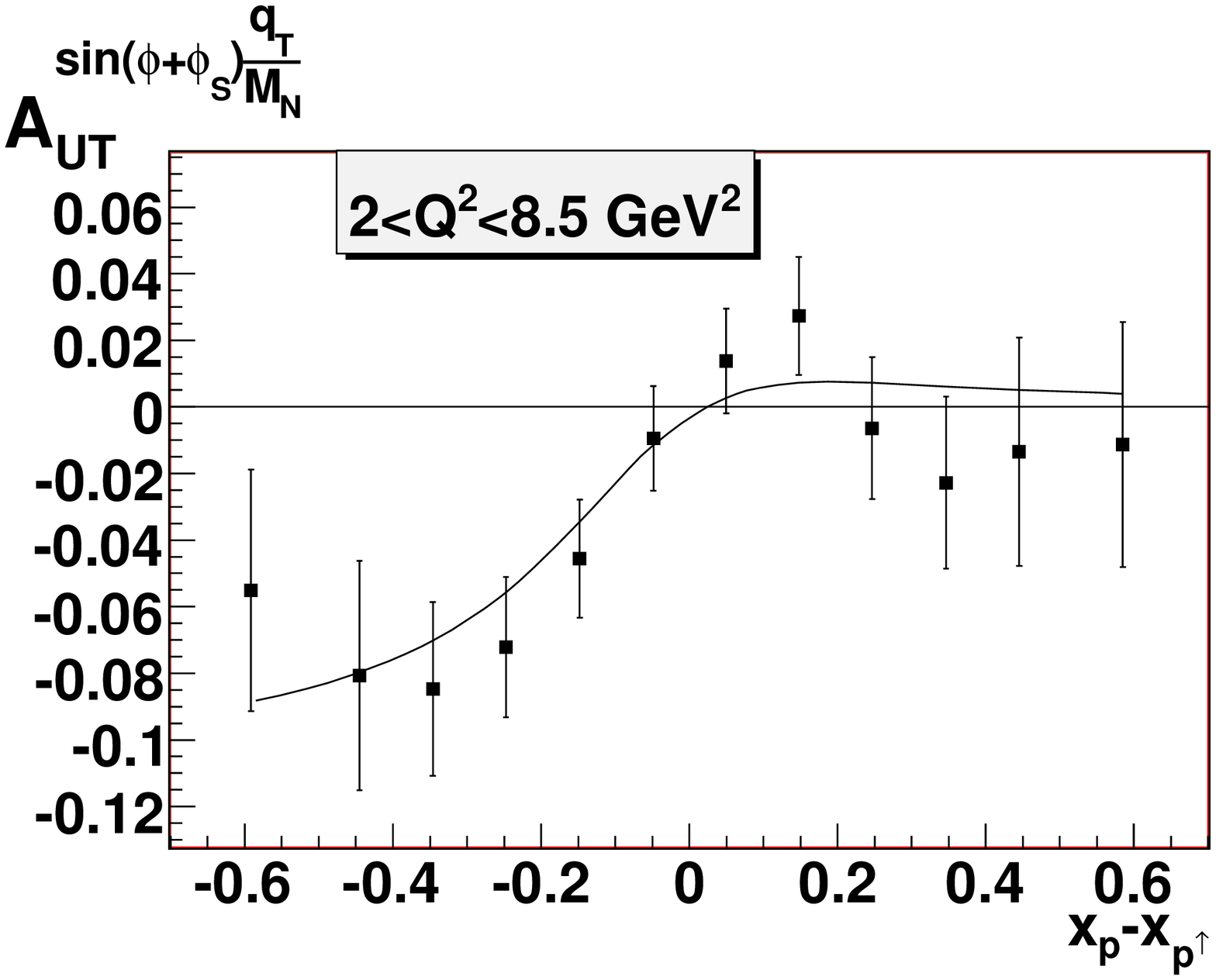} & 
\includegraphics[height=5cm]{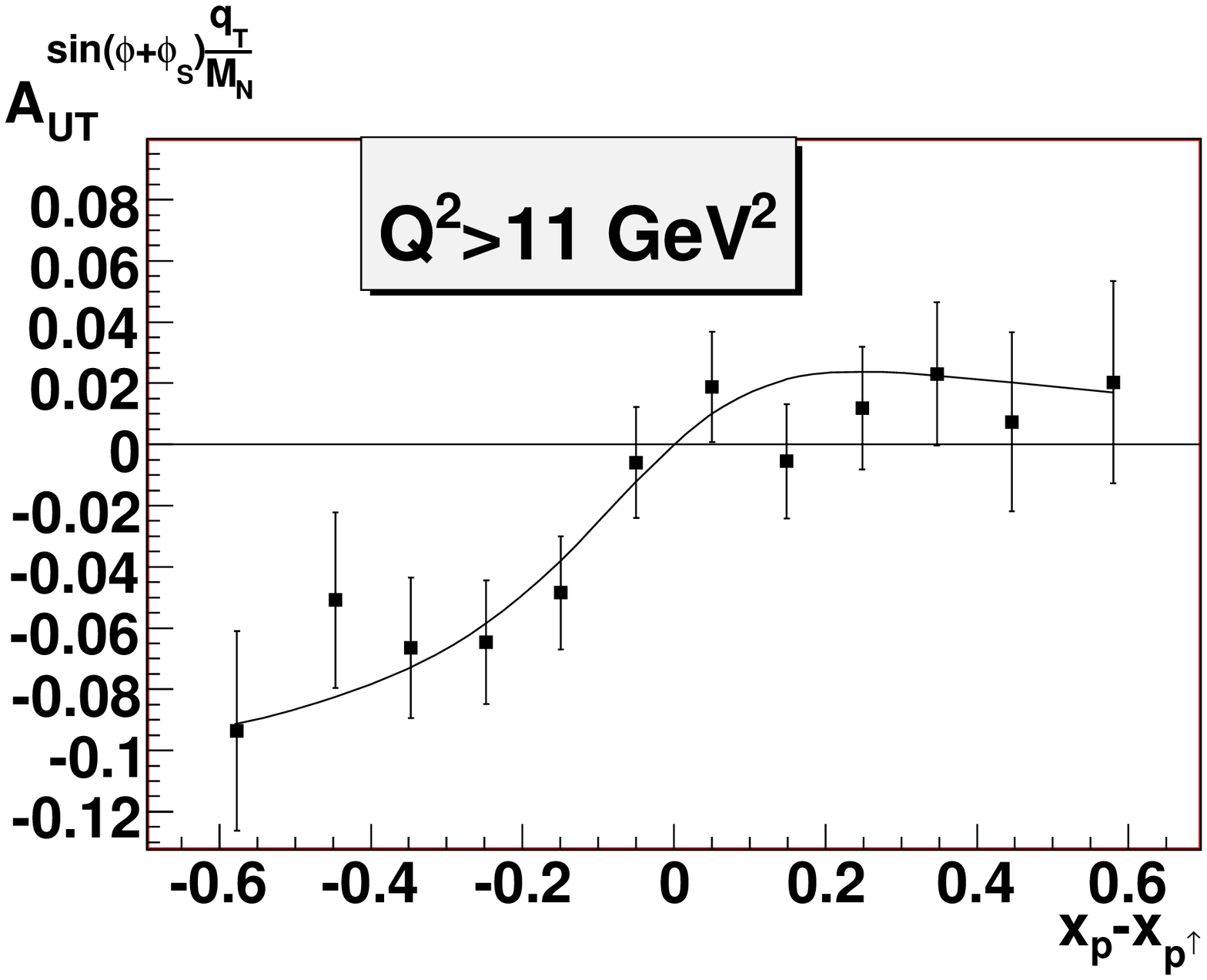}\\
\includegraphics[height=5cm]{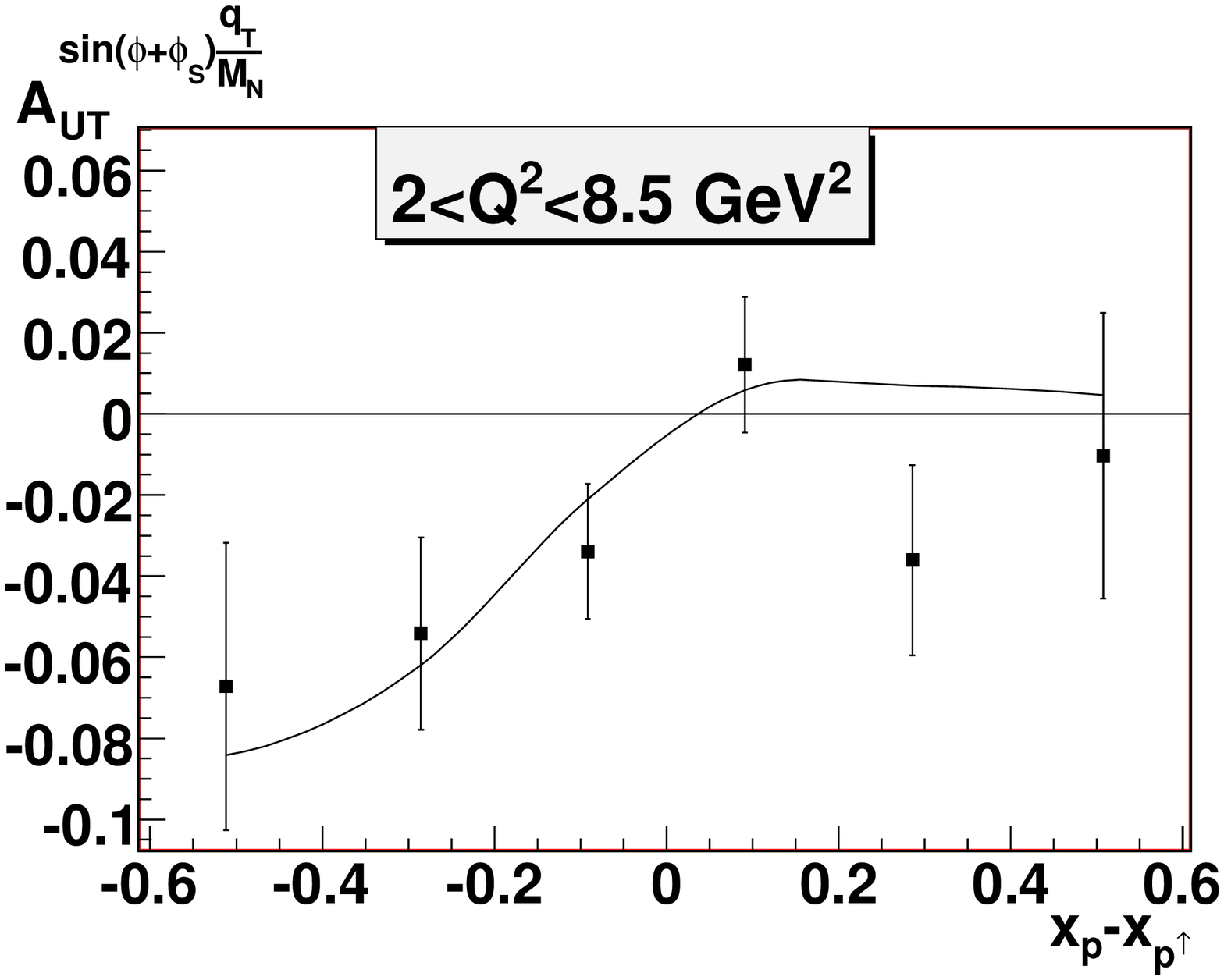} & 
\includegraphics[height=5cm]{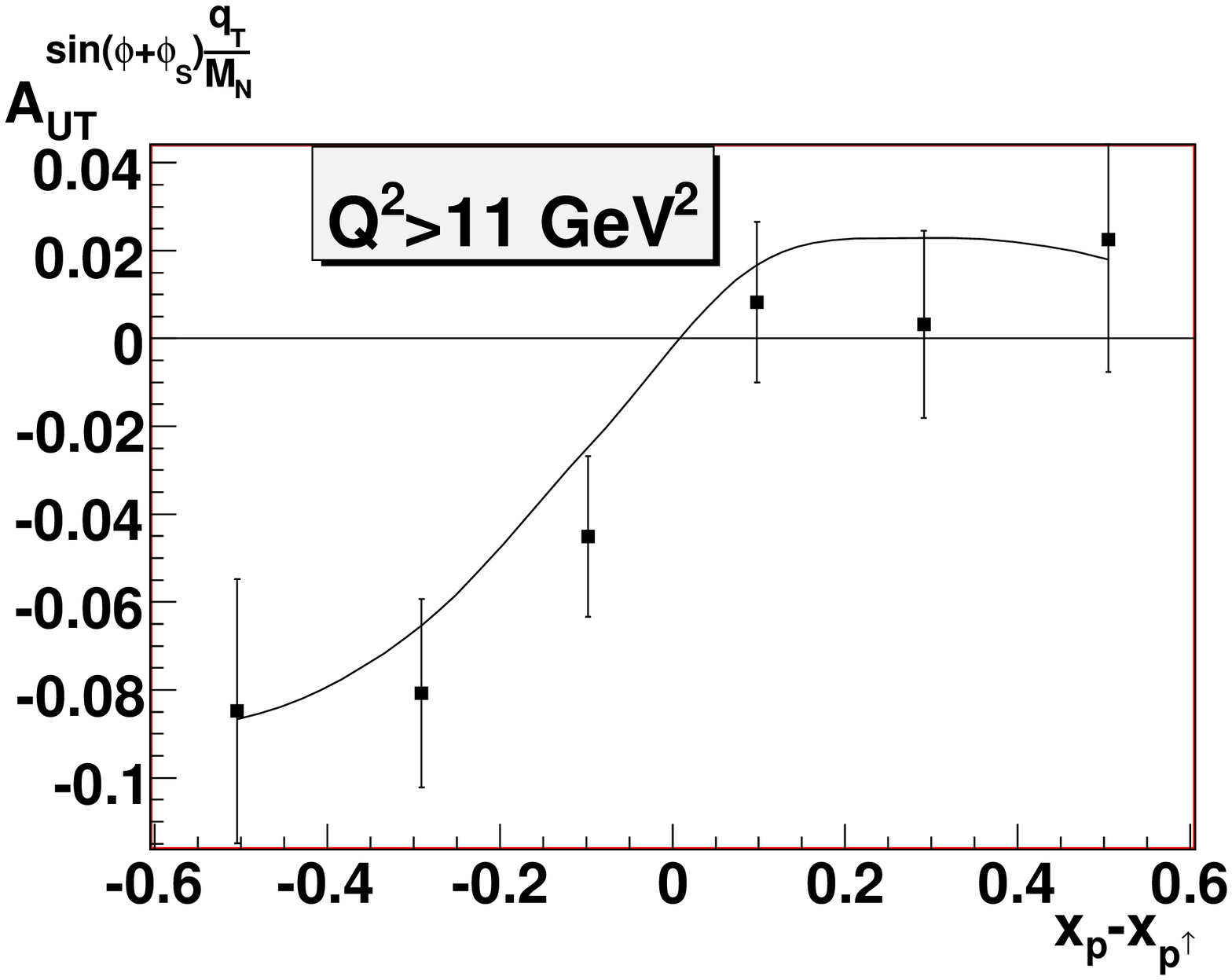}

        \end{tabular}
        \caption{Estimation of asymmetries $A_{UT}^{\sin(\phi+\phi_S)\frac{q_T}{M_N}}\palka_{pp^\uparrow}$  for NICA, $s=400GeV^2$. 
        Here the evolution model with the input ansatz $h_{1q,\bar q}=\Delta q,\bar q$
        at $Q_0^2=0.23\,GeV^2$ is used.
GRV94 \cite{grv94} parametrization for $q(x)$ and GRSV95 \cite{GRSV95} parametrization for $\Delta q(x)$ are used.
The points with error bars are obtained with the developed generator of polarized DY events at the applied
statistics 100K (top) and 50K (bottom) pure Drell-Yan events. 
$\langle Q^2\rangle\simeq3.5\,GeV^2$ and $\langle Q^2\rangle=15\,GeV^2$ for the left and right plots, respectively. 
}
\label{fig:simulations_nica_100k}
\end{figure}

\begin{figure}
        \begin{tabular}{cc}
\includegraphics[height=5cm]{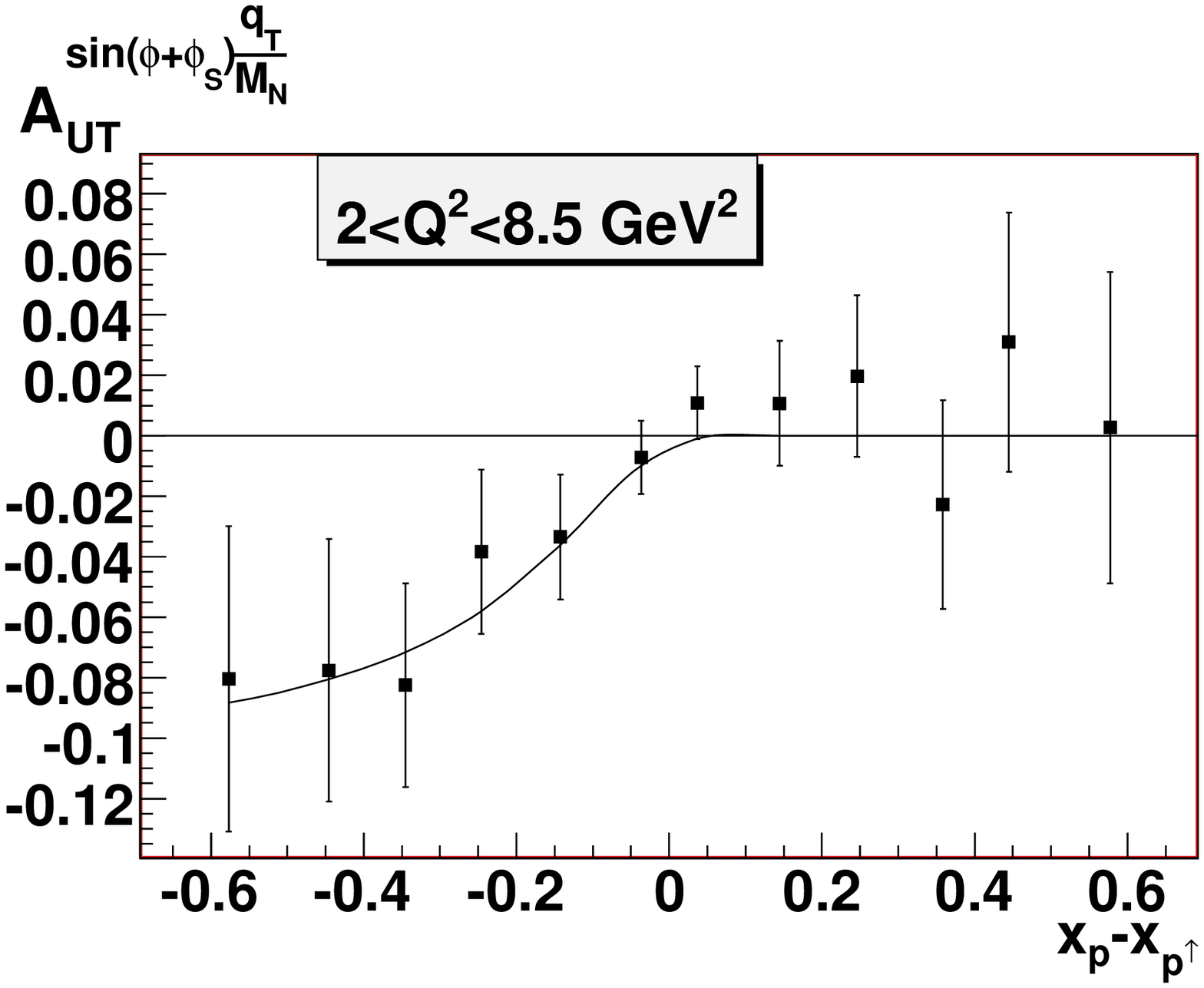} & 
\includegraphics[height=5cm]{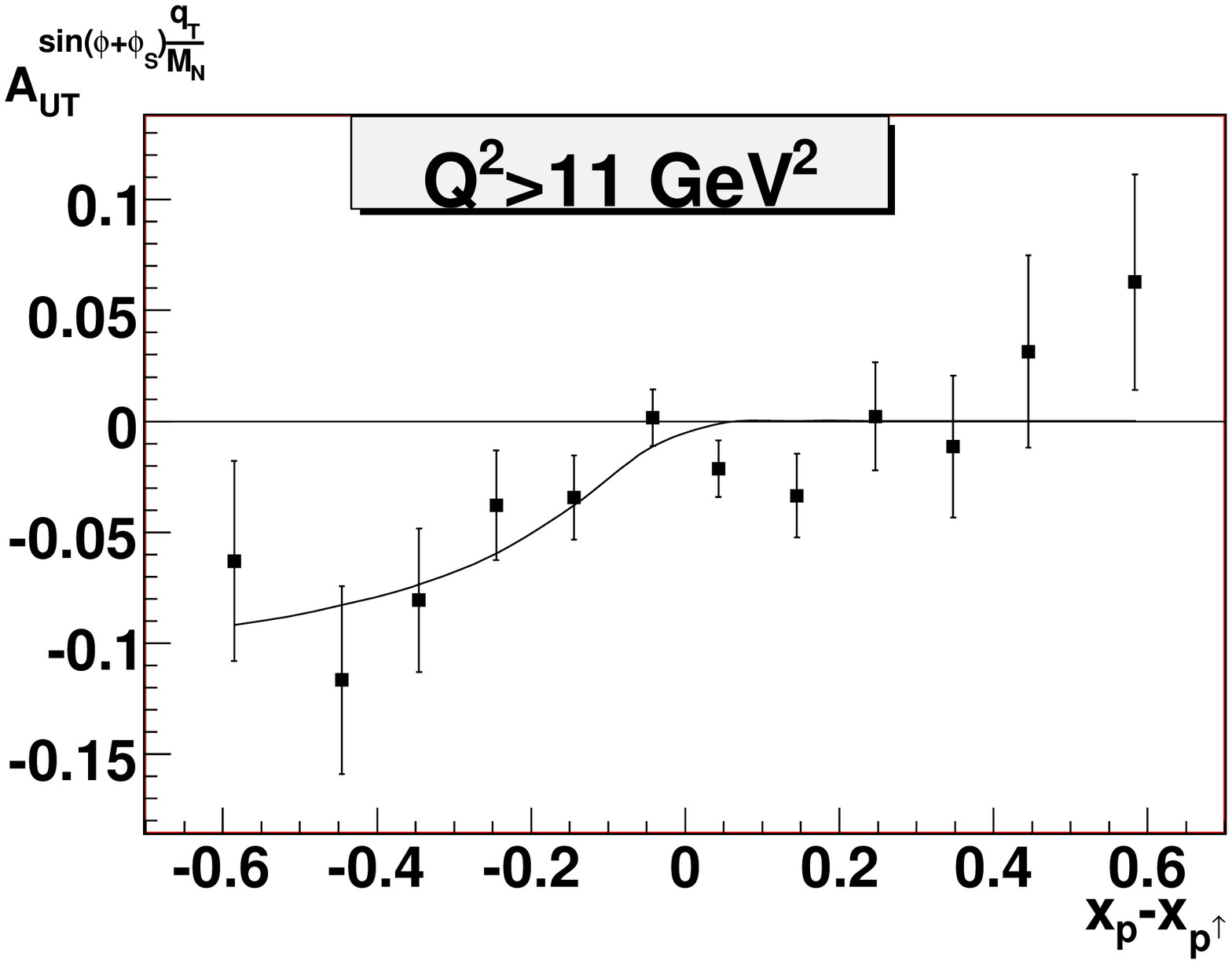}\\
\includegraphics[height=5cm]{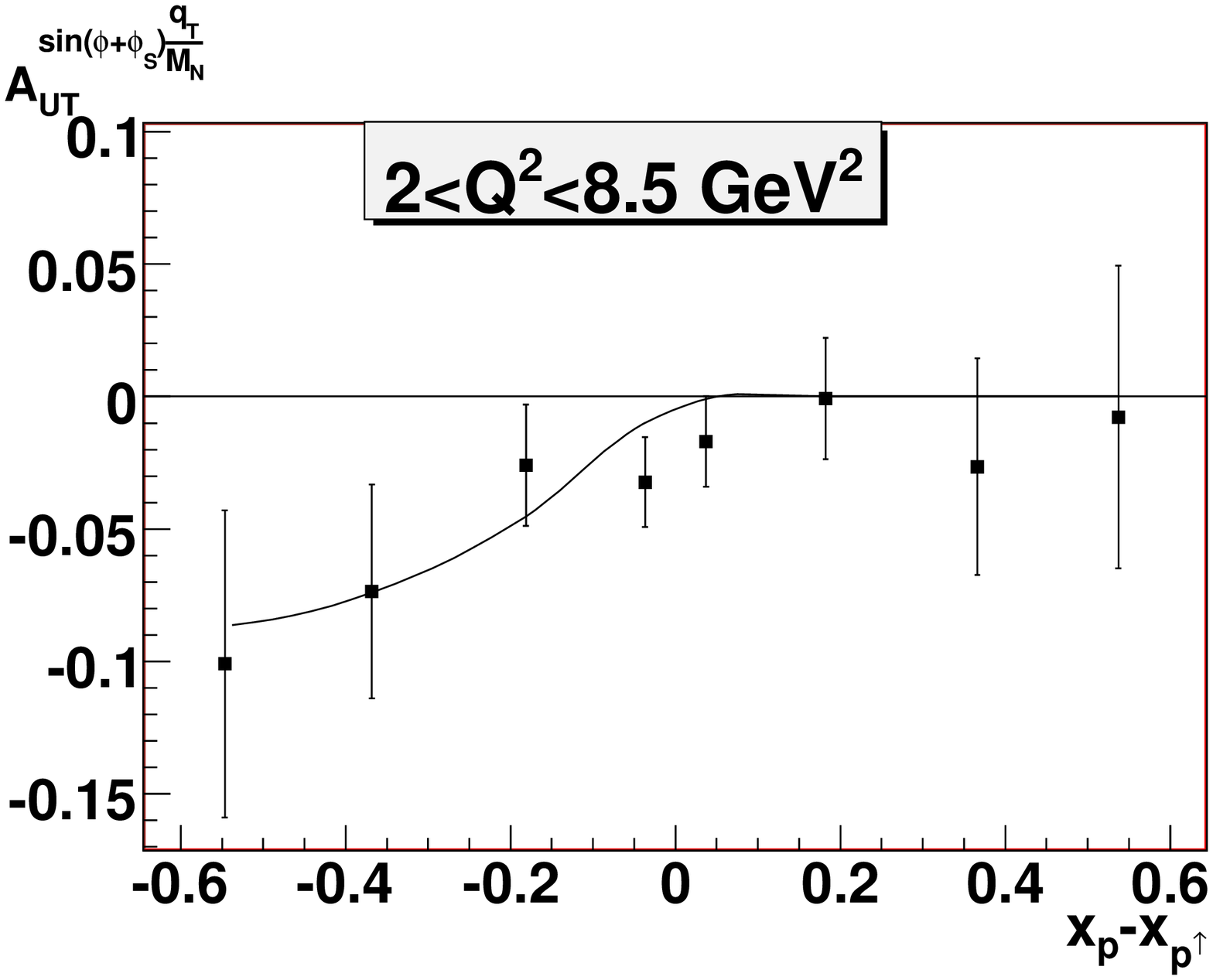} & 
\includegraphics[height=5cm]{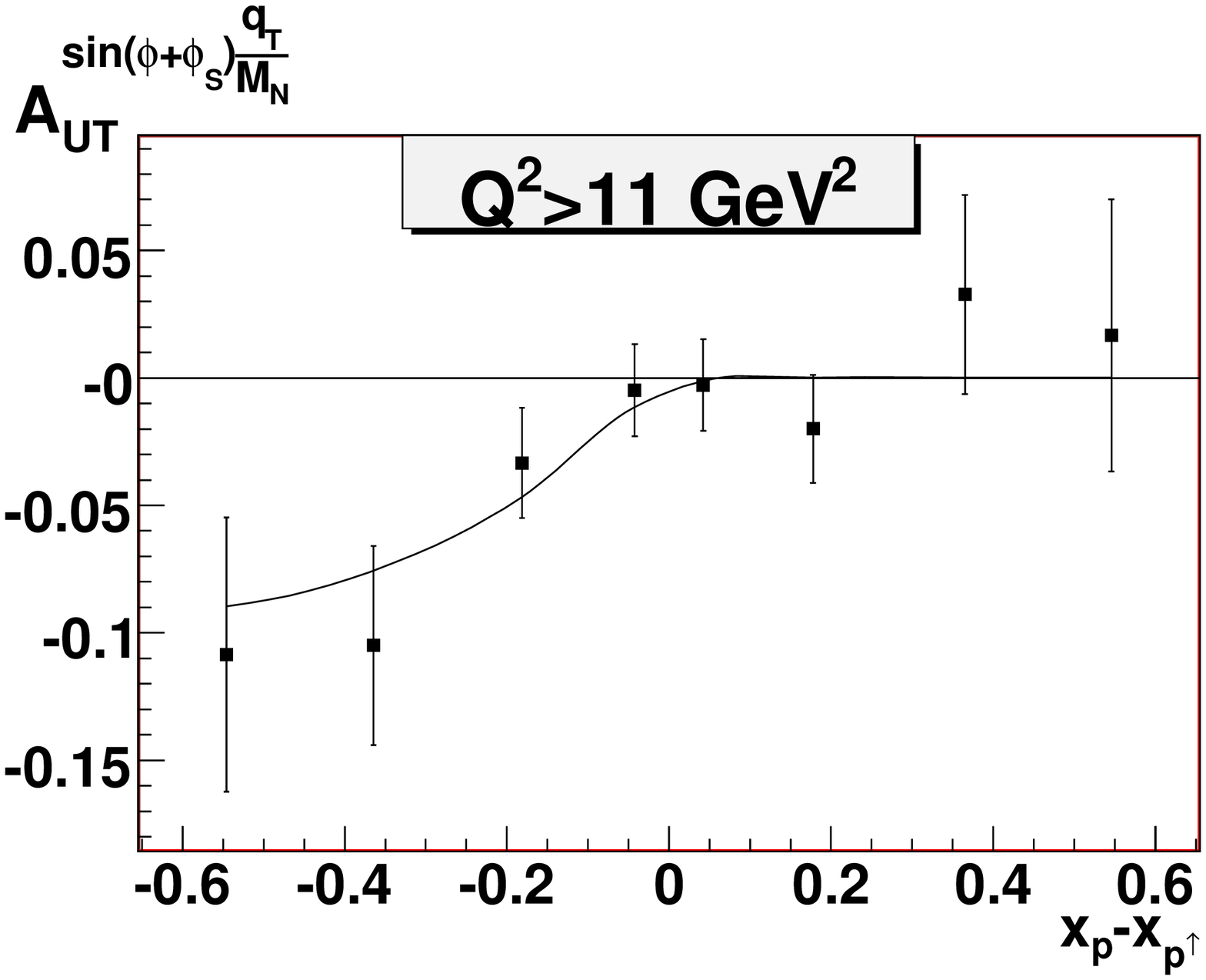}

        \end{tabular}
        \caption{Estimation of asymmetries $A_{UT}^{\sin(\phi+\phi_S)\frac{q_T}{M_N}}\palka_{pp^\uparrow}$  for RHIC, $s=200^2GeV^2$. 
        Here the evolution model with the input ansatz $h_{1q,\bar q}=\Delta q,\bar q$
        at $Q_0^2=0.23\,GeV^2$ is used.
GRV94 \cite{grv94} parametrization for $q(x)$ and GRSV95 \cite{GRSV95} parametrization for $\Delta q(x)$ are used.
The points with error bars are obtained  with the developed generator of polarized DY events at the applied
statistics 100K (top) and 50K (bottom) pure Drell-Yan events.
$\langle Q^2\rangle\simeq3.9\,GeV^2$ and $\langle Q^2\rangle=22\,GeV^2$ for the left and right plots, respectively. 
}
\label{fig:simulations_rhic_100k}
\end{figure}

Let us now estimate the feasibility of SSA  $A_{UT}^{\sin(\phi+\phi_S)\frac{q_T}{M_N}}$
giving us an access to transversity and Boer-Mulders PDFs.
We again (see Section 3) present the estimations for $pp^\uparrow$ collisions and for two quite different 
RHIC and NICA center of mass energies.
The results are  presented in Figs. \ref{fig:simulations_nica_100k} 
and \ref{fig:simulations_rhic_100k}. 
 For COMPASS and J-PARC kinematics our calculations
produce almost the same figures as for NICA center of mass energy, so that we again omit the respective plots.
For the simulations with the developed generator 
we again use (see Section 3) Boer model for $h_1^{\perp}$
and the evolution model for $h_1$ with $h_{1q(\bar q)}=\Delta q(\Delta\bar q)$
ansatz at initial scale $Q_0^2=0.23\,GeV^2$. 

Looking at Figs. \ref{fig:simulations_nica_100k} 
and \ref{fig:simulations_rhic_100k} one can see that in the region
$x_p<x_{p^\uparrow}$ even at 
statistics 50K pure Drell-Yan events (bottom part of   Fig. \ref{fig:simulations_nica_100k})
one can hope to see within the errors at least 
three points for $A_{UT}^{\sin(\phi+\phi_S)\frac{q_T}{M_N}}$  
in the kinematical region $x_p-x_{p^\uparrow}<0$. 
At the same time, at the statistics 100K events one can hope also to reconstruct the functional form 
of this SSA (see top part of Fig. \ref{fig:simulations_nica_100k})
in the kinematical region $x_p<x_{p^\uparrow}$. Regretfully, the kinematical region $x_p<x_{p^\uparrow}$
is hardly accessible for COMPASS because of the forward geometry spectrometer and the unpolarized
proton beam available to this running experiment. In the case of another experiment with
the fixed target,  planning at J-PARC facility, one could reach this kinematical region if the polarized proton beam
would be available (this option is now planned at J-PARC \cite{sawada}). Fortunately, there are no any problems
to reach the kinematical region $x_p<x_{p^\uparrow}$ for the symmetric collider mode available to RHIC and
NICA.

Let us note that due to the close values of $A_{UT}^{\sin(\phi+\phi_S)\frac{q_T}{M_N}}$ 
in the cases of $pp^\uparrow$ and $Dp^\uparrow$ collisions (see Fig. \ref{fig:dp_bm}) all conclusions
concerning feasibility of  $A_{UT}^{\sin(\phi+\phi_S)\frac{q_T}{M_N}}$ in the case of $pp^\uparrow$ collisions 
are valid in $Dp^\uparrow$ case too.

Looking at Figs. \ref{fig:simulations_sivers_100k}-\ref{fig:simulations_rhic_100k}
one can see that if the available experimental statistics
will be rather low (less than 50K events) 
it is hardly possible to reconstruct the functional form of Sivers and Boer-Mulders PDFs.
However, even in this unfavorable case one can hope at least to check very important QCD prediction
(\ref{sign}).

\begin{figure}
        \begin{tabular}{cc}
\includegraphics[height=5cm]{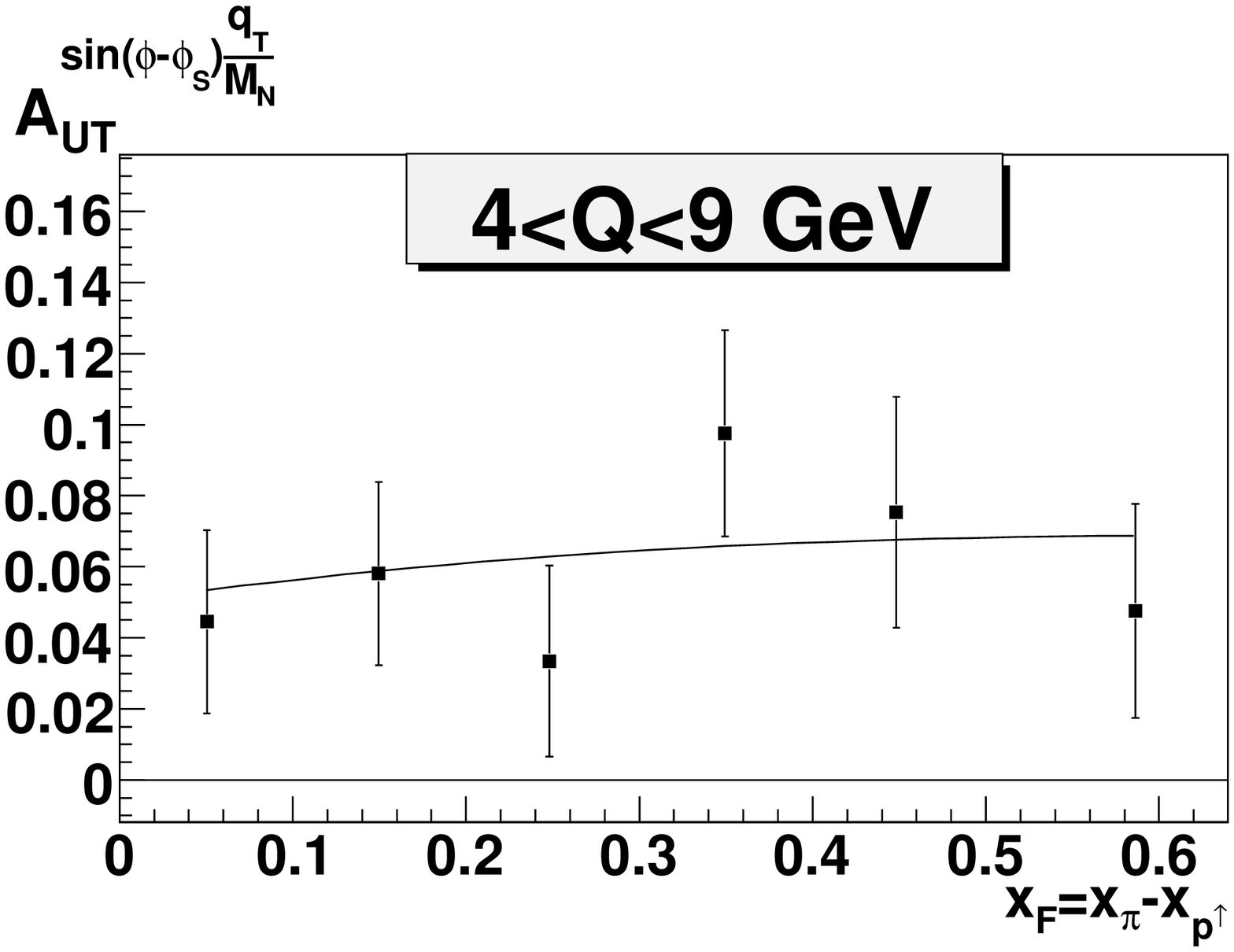} & 
\includegraphics[height=5cm]{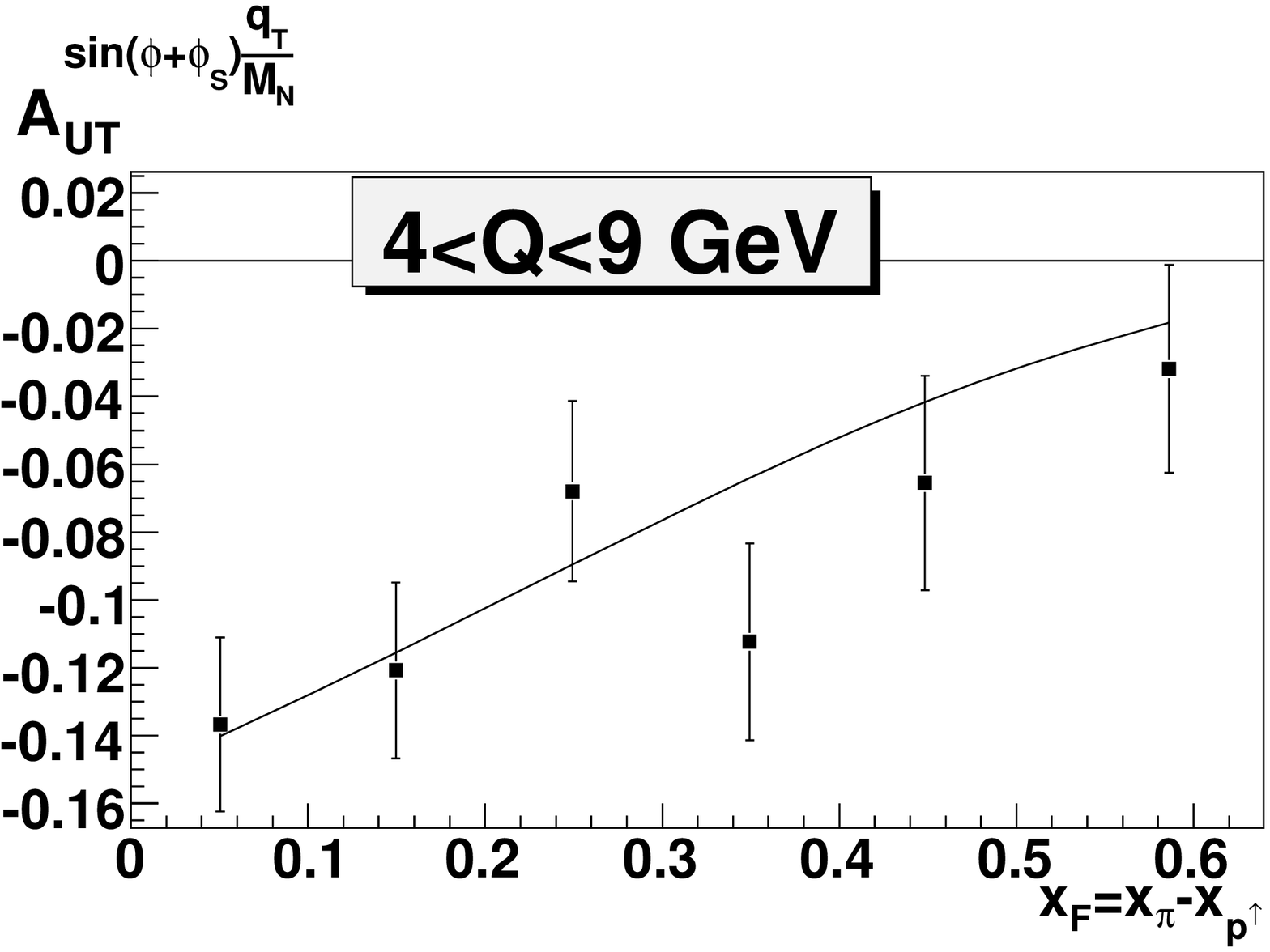}
        \end{tabular}
        \caption{Estimations on SSA $A_{UT}^{\sin(\phi-\phi_S)\frac{q_T}{M_N}}\palka_{\pi^-p^\uparrow}$ (left) 
        and $A_{UT}^{\sin(\phi+\phi_S)\frac{q_T}{M_N}}\palka_{\pi^-p^\uparrow}$ (right) feasibility for COMPASS,
        $s=300\,GeV^2$. 
        Magnitude of $A_{UT}^{\sin(\phi+\phi_S)\frac{q_T}{M_N}}\palka_{\pi^-p^\uparrow}$ (solid curve)
        is estimated just as in 
        Ref. \cite{our2} (see discussion on Figs. 3 and 4 in Ref. \cite{our2}).
        For estimation of $A_{UT}^{\sin(\phi-\phi_S)\frac{q_T}{M_N}}\palka_{\pi^-p^\uparrow}$ (solid curve)
        fit from \cite{efremov_new} for $f_{1T}^{\perp(1)q}\palka_{p^\uparrow}$ and parametrization \cite{grvp} on $f_{1q}\palka_{\pi^-}$ are used.
The points with error bars are obtained  with the developed generator of polarized DY events at the applied
statistics  50K  pure Drell-Yan events.
$\langle Q^2\rangle\simeq25\,GeV^2$ for both plots. 
}
\label{fig:simulations_compass}
\end{figure}

In conclusion of this section let us estimate the feasibility of both 
$A_{UT}^{\sin(\phi-\phi_S)\frac{q_T}{M_N}}$ and $A_{UT}^{\sin(\phi+\phi_S)\frac{q_T}{M_N}}$ SSA
for the $\pi^-p^\uparrow\to\mu^+\mu^-X$  DY processes available to COMPASS. 
The model estimation of SSA $A_{UT}^{\sin(\phi+\phi_S)\frac{q_T}{M_N}}$ magnitude was already presented
in the previous paper \cite{our2}. However, that time we could not present the feasibility estimations
for this SSA because there still was not any generator of polarized DY events
in the free access. Besides, only recently the optimal kinematical conditions 
for DY program at COMPASS were chosen. These are the pion beam energy
$160$ GeV  and the range $4<Q<9$ GeV for the invariant dilepton mass \cite{compass_dy}. 
The results on SSA $A_{UT}^{\sin(\phi\pm\phi_S)\frac{q_T}{M_N}}\palka_{\pi^-p^\uparrow}$  
are presented in Fig. \ref{fig:simulations_compass}. Because of the restriction (\ref{restriction})
for the COMPASS forward geometry spectrometer we present the results only for 
$x_\pi>x_{p^\uparrow}$. Looking at Fig. \ref{fig:simulations_compass} one can see that 
even at rather low applied statistics 50K events there six presumably measurable points for both asymmetries,
which encourage us that it could be possible to reconstruct the functional form for both SSA. 
Thus, one can conclude that for COMPASS it is much more profitable to study DY processes
with the pion-proton collisions. At the same time, as it was discussed above, DY processes 
$p(D)p^\uparrow(D^\uparrow)\to l^+l^-X$ could be most successfully studied in collider mode 
at RHIC and NICA.

\section{Conclusion}
{\it In summary}, the Drell-Yan processes with the colliding protons and deutrons 
available to RHIC, NICA, COMPASS and J-PARC were considered. 
We estimated the   single-spin asymmetries
$A_{UT}^{\sin(\phi-\phi_S)\frac{q_T}{M_N}}$ and $A_{UT}^{\sin(\phi+\phi_S)\frac{q_T}{M_N}}$, which
give us an access to Sivers and to Boer-Mulders and transversity PDFs, respectively.
The estimations were performed for the different $\sqrt{s}$ values, corresponding to 
the kinematical conditions of  RHIC, NICA, J-PARC and COMPASS facilities.
The preliminary estimations for $pp^\uparrow$ collisions demonstrate that 
SSA $A_{UT}^{\sin(\phi-\phi_S)\frac{q_T}{M_N}}$ can reach quite considerable values (5-10\%) 
in both $x_p> x_{p^\uparrow}$ and  $x_p< x_{p^\uparrow}$ regions.
On other hand, the estimations performed for SSA $A_{UT}^{\sin(\phi+\phi_S)\frac{q_T}{M_N}}$ 
show that this asymmetry is negligible in the region $x_p> x_{p^\uparrow}$ and takes considerable values (also about
5-10\%) in the region  $x_p< x_{p^\uparrow}$.
The estimations performed for SSA in the case of $Dp$ and $DD$ collisions demonstrate that 
the asymmetries for DY processes with $pD^\uparrow$ collisions are compatible with zero except for perhaps one, containing 
sum $h_{1u}+h_{1d}$. The latest analysis \cite{ansel_kotz} predicts that this sum could essentially differ from
zero. Certainly, because of very large uncertainties, this is very preliminary conclusion which should be carefully checked
in the future DY experiments. 
On the contrary to DY processes with $pD^\uparrow$ and $DD^\uparrow$ collisions, 
SSA for $Dp^\uparrow$ collisions are close in their values to the respective SSA for $pp^\uparrow$ collisions
and, thus, presumably could be feasible in the same kinematical regions. While SSA 
$A_{UT}^{\sin(\phi-\phi_S)\frac{q_T}{M_N}}\palka_{Dp^\uparrow}$ is about 50-80\% of
$A_{UT}^{\sin(\phi-\phi_S)\frac{q_T}{M_N}}\palka_{pp^\uparrow}$, the asymmetry 
$A_{UT}^{\sin(\phi+\phi_S)\frac{q_T}{M_N}}\palka_{Dp^\uparrow}$ is even slightly larger 
than $A_{UT}^{\sin(\phi+\phi_S)\frac{q_T}{M_N}}\palka_{pp^\uparrow}$.

The new generator of polarized DY events was developed which allowed us 
to estimate the feasibility of both weighted with $\sin(\phi-\phi_S)$ and  $\sin(\phi+\phi_S)$ single-spin asymmetries. 
These estimations performed for proton-proton collisions demonstrate  
that  both SSA are presumably measurable even at the applied statistics 50K pure Drell-Yan events.
While $A_{UT}^{\sin(\phi-\phi_S)\frac{q_T}{M_N}}$ is presumably measurable in both
kinematical regions $x_p>x_{p^\uparrow}$ and $x_p<x_{p^\uparrow}$, the asymmetry
$A_{UT}^{\sin(\phi+\phi_S)\frac{q_T}{M_N}}$ could be measured only in the region 
$x_p<x_{p^\uparrow}$, where it takes quite considerable values.
It is of importance that while for the symmetric collider mode (RHIC, NICA) there is no
any difficulties to reach both $x_p>x_{p^\uparrow}$ and $x_p<x_{p^\uparrow}$ kinematical regions, 
for the fixed target mode (COMPASS, J-PARC), 
there is the problem to reach the region $x_p<x_{p^\uparrow}$. This is hardly possible 
for the running COMPASS experiment, where only unpolarized proton beam and forward geometry spectrometer
is available. At the same time it still could be achieved by J-PARC, where at present the option
with polarized proton beam is planned.

We also studied the behavior of SSA in two different 
limiting cases, $x_p\gg x_{p^\uparrow}$ and $x_p\ll x_{p^\uparrow}$. It is of importance
that in these  limiting cases  one can 
 essentially reduce the number of unknown PDFs entering the asymmetries. In particular,  studying 
the unpolarized and single-polarized Drell-Yan processes in the limiting case $x_p\ll x_{p^\uparrow}$
one can directly extract the ratio of transversity and first moment of Boer-Mulders PDF.

Thus, one can conclude that it is much better to study the single-polarized DY 
processes with $pp$, $pD$ and $DD$ collisions
in the symmetric collider mode (RHIC and NICA). For fixed target experiments (like J-PARC and COMPASS)
the polarized proton beam is necessary to improve the situation. This option is planned 
at J-PARC but is hardly possible for already running COMPASS experiment, 
where it is much more profitable to study DY processes
with the pion-proton collisions (namely this option now is planned at COMPASS \cite{compass_dy}).

Let us stress once again that any new measurements of SSA in Drell-Yan processes 
are of extreme importance even at rather poor available statistics
of collected Drell-Yan events. 
Even in this unfavorable case  
one can at least to check very important QCD prediction (\ref{sign}),
which would provide a crucial test of our understanding of T-odd effects
within QCD and the factorization approach to the processes sensitive to transverse parton momenta. 
It is encouraging that today the Drell-Yan measurements are planned simultaneously 
in the different world centers for high energy physics.

\begin{center}
Acknowledgments
\end{center}
  The authors are grateful to R.~Bertini,
 O.~Denisov,  A.~Efremov, Y.~Goto, T.~Iwata, S.~Kazutaka, V.~Kekelidze, S.~Kumano, 
  A.~Maggiora, I.~Meshkov,  A.~Olshevsky, 
G.~Piragino, G.~Pontecorvo, S.~Sawada, I.~Savin, A.~Sorin and O.~Teryaev,
 for fruitful discussions.

The work of O. Shevchenko, A. Nagaytsev and O. Ivanov was supported by the Russian Foundation
for Basic Research (Project No. 07-02-01046).


\end{document}